\documentclass{article}
\input epsf.sty

\def\lya{Ly$\alpha$~}
\def\Lya{Ly$\alpha$~}
\def\HI{\rm H\,I~}
\def\zr{z_{\rm reion}}
\def\beq{\begin{equation}}
\def\beqa{\begin{eqnarray}}
\def\eeq{\end{equation}}
\def\eeqa{\end{eqnarray}}
\def\Omm{{\Omega_m}}
\def\gtrsim{>}
\def\ga{>}
\def\lesssim{<}
\def\la{{\lesssim}}
\def\Ommz{{\Omega_m^{\,z}}}
\def\Omr{{\Omega_r}}
\def\Omk{{\Omega_k}}
\def\Oml{{\Omega_{\Lambda}}}
\def\sun{\odot}
\def\xb{{\bf x}}
\def\rb{{\bf r}}
\def\vb{{\bf v}}
\def\ub{{\bf u}}
\def\kb{{\bf k}}
\def\kB{k}
\def\cN{c_{\rm N}}

\begin{document}
\hfill {\it UNESCO EOLSS ENCYCLOPEDIA}\bigskip\bigskip 

\centerline
{\it{\Large ``Let
there be Light'': the Emergence of Structure}}
\centerline{\it{\Large out of the Dark Ages in the
Early Universe}}\bigskip\bigskip

\def\Ref#1{\noindent\hangindent=30pt\hangafter=1\nobreak
\frenchspacing #1 \par}
\def\sm{\smallskip}

{\large
{\bf Abraham Loeb}

Department of Astronomy,
Harvard University, 60 Garden St., Cambridge MA, 02138
}


\section*{Contents}

\noindent
1. Introduction

1.1 Observing our past

1.2 The expanding universe

\noindent
2. Galaxy Formation

2.1. Growth of linear perturbations

2.2. Halo properties

2.3. Formation of the first stars

2.4. Gamma-ray Bursts: probing the first stars one star at a time

2.5. Supermassive black holes

2.6. The epoch of reionization

2.7. Post-reionization suppression of low-mass galaxies

\noindent
3. Probing the Diffuse Intergalactic Hydrogen

3.1. Lyman-alpha absorption

3.2. 21-cm absorption or emission

\noindent
4. Conclusions

\section*{Summary}
Cosmology is by now a mature experimental science.  We are privileged
to live at a time when the story of genesis (how the Universe started and
developed) can be critically explored by direct observations.  Looking deep
into the Universe through powerful telescopes we can see images of the
Universe when it was younger, because of the finite time it takes light to
travel to us from distant sources.

Existing data sets include an image of the Universe when it was 0.4 million
years old (in the form of the cosmic microwave background), as well as
images of individual galaxies when the Universe was older than a billion
years.  But there is a serious challenge: in between these two epochs was a
period when the Universe was dark, stars had not yet formed, and the cosmic
microwave background no longer traced the distribution of matter.  And this
is precisely the most interesting period, when the primordial soup evolved
into the rich zoo of objects we now see.  


The observers are moving ahead along several fronts.  The first involves
the construction of large infrared telescopes on the ground and in space,
that will provide us with new photos of the first galaxies.  Current plans
include ground-based telescopes which are 24-42 meter in diameter, and
NASA's successor to the Hubble Space Telescope, called the James Webb Space
Telescope.  In addition, several observational groups around the globe are
constructing radio arrays that will be capable of mapping the
three-dimensional distribution of cosmic hydrogen in the infant
Universe. These arrays are aiming to detect the long-wavelength (redshifted
21-cm) radio emission from hydrogen atoms.
The images from these antenna arrays will reveal how the non-uniform
distribution of neutral hydrogen evolved with cosmic time and eventually
was extinguished by the ultra-violet radiation from the first galaxies.
Theoretical research has focused in recent years on predicting the expected
signals for the above instruments and motivating these ambitious
observational projects.

\section{Introduction}

\subsection{Observing our past}

\begin{figure}
\epsfxsize=15cm \epsfbox{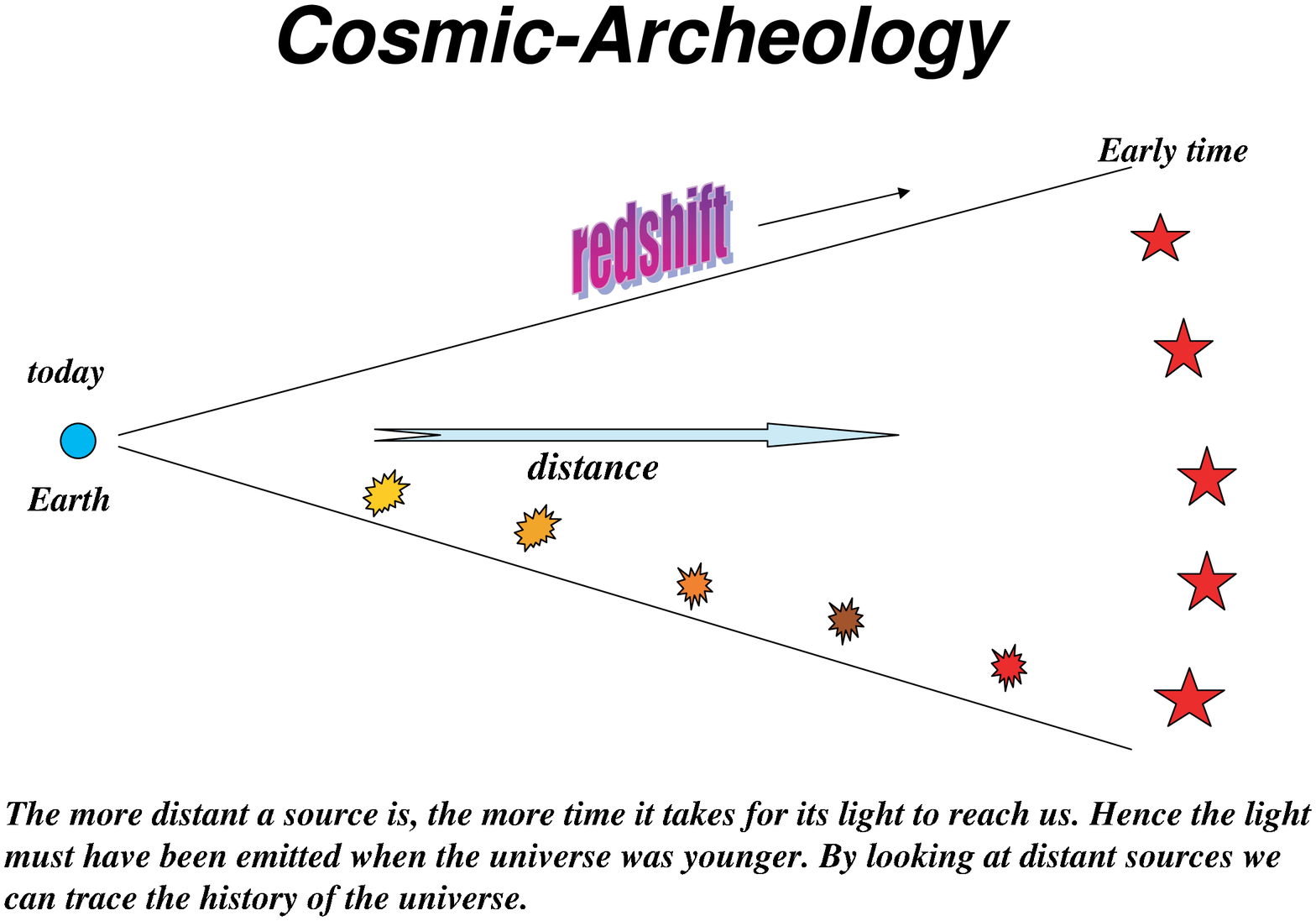}
\caption{Cosmology is like archeology. The deeper one looks, the older is
the layer that is revealed, owing to the finite propagation speed of light.
}
\label{fig:z}
\end{figure}

When we look at our image reflected off a mirror at a distance of 1
meter, we see the way we looked 6.7 nanoseconds ago, the light travel
time to the mirror and back. If the mirror is spaced $10^{19}~{\rm cm}
\simeq 3~$pc away, we will see the way we looked twenty one years
ago. Light propagates at a finite speed, and so by observing distant
regions, we are able to see what the Universe looked like in the past,
a light travel time ago (Figure~\ref{fig:z}). The statistical
homogeneity of the Universe on large scales guarantees that what we
see far away is a fair statistical representation of the conditions
that were present in in our region of the Universe a long time ago.

\begin{figure}
\epsfxsize=10cm \epsfbox{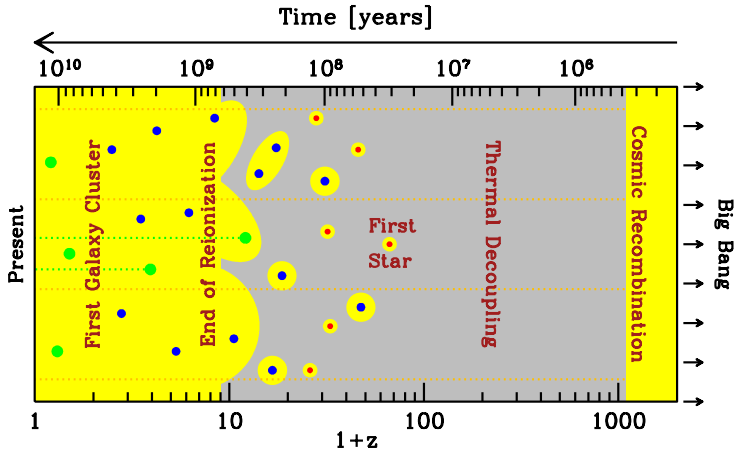}
\caption{Overview of cosmic history, with the age of the universe shown on
the top axis and the corresponding redshift on the bottom axis. Yellow
represents regions where the hydrogen is ionized, and gray, neutral
regions. Stars form in galaxies located within dark matter concentrations
whose typical mass grows with time, starting with $\sim 10^5 M_{\odot}$
(red circles) for the host of the first star, rising to $10^7$--$10^9
M_{\odot}$ (blue circles) for the sources of reionization, and reaching
$\sim 10^{12} M_{\odot}$ (green circles) for present-day galaxies like our
own Milky Way. Astronomers probe the evolution of the cosmic gas using the
absorption of background light (dotted lines) by atomic hydrogen along the
line of sight. The classical technique uses absorption by the
Lyman-$\alpha$ resonance of hydrogen of the light from bright quasars
located within massive galaxies, while a new type of astronomical
observation will use the 21-cm line of hydrogen with the cosmic microwave
background as the background source.}
\label{fig:history}
\end{figure}

This fortunate situation makes cosmology an empirical science. We do
not need to guess how the Universe evolved. Using telescopes we can
simply see how it appeared at earlier cosmic times. In principle, this
allows the entire 13.7 billion year cosmic history of our universe to
be reconstructed by surveying the galaxies and other sources of light
to large distances (Figure~\ref{fig:history}). Since a greater
distance means a fainter flux from a source of a fixed luminosity, the
observation of the earliest sources of light requires the development
of sensitive instruments and poses challenges to observers.

As the universe expands, photon wavelengths get stretched as well.
The factor by which the observed wavelength is increased (i.e. shifted
towards the red) relative to the emitted one is denoted by $(1+z)$, where
$z$ is the cosmological redshift.  Astronomers use the known emission
patterns of hydrogen and other chemical elements in the spectrum of each
galaxy to measure $z$.  This then implies that the universe has expanded by
a factor of $(1+z)$ in linear dimension since the galaxy emitted the
observed light, and cosmologists can calculate the corresponding distance
and cosmic age for the source galaxy. Large telescopes have allowed
astronomers to observe faint galaxies that are so far away that we see them
more than twelve billion years back in time. Thus, we know directly that
galaxies were in existence as early as 850 million years after the Big
Bang, at a redshift of $z \sim 6.5$ or higher.

\begin{figure}
\epsfxsize=10cm \epsfbox{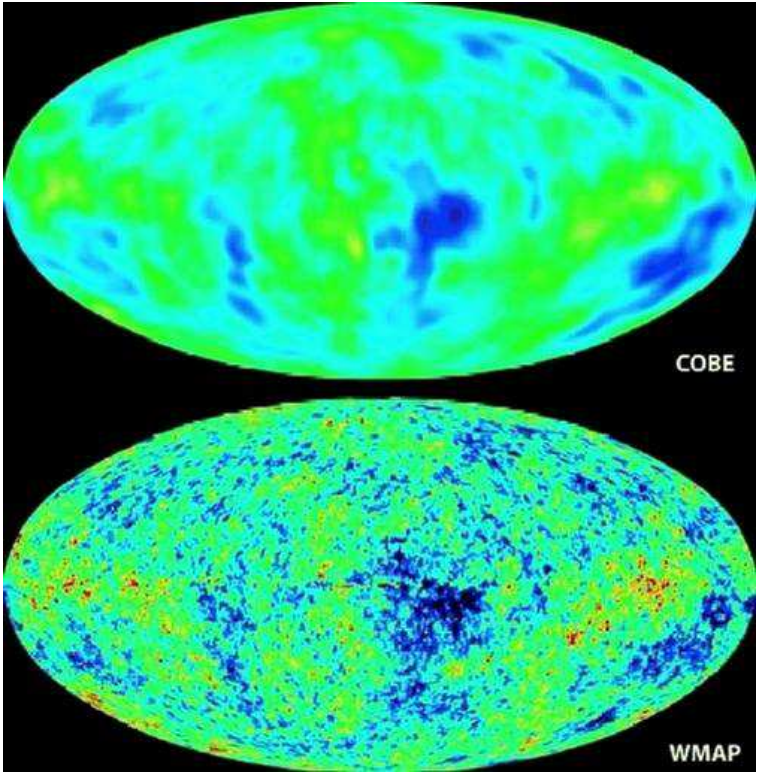}
\caption{Images of the Universe shortly after it became transparent, taken
by the {\it COBE}\/ and {\it WMAP}\/ satellites (see
http://map.gsfc.nasa.gov/ for details). The slight density inhomogeneities
in the otherwise uniform Universe imprinted a map of hot and cold spots
(shown here as different colors) in the CMB that is observed today. The
existence of these anisotropies was predicted three decades before the
technology for taking these images became available, in a number of
theoretical papers including Sachs \& Wolfe (1967), Rees \& Sciama (1968),
Silk (1968), Sunyaev \& Zeldovich (1970), and Peebles \& Yu (1970).}
\label{fig:CMB}
\end{figure}

We can in principle image the Universe only if it is transparent. Earlier
than $400\,000$ years after the big bang, the cosmic hydrogen was broken
into its constituent electrons and protons (i.e. ``ionized'') and the
Universe was opaque to scattering by the free electrons in the dense
plasma. Thus, telescopes cannot be used to electromagnetically image the
infant Universe at earlier times (or redshifts $\ga 10^3$). The earliest
possible image of the Universe was recorded by the COBE and WMAP
satellites, which measured the temperature distribution of the cosmic
microwave background (CMB) on the sky (Figure~\ref{fig:CMB}).

The CMB, the relic radiation from the hot, dense beginning of the universe,
is indeed another major probe of observational cosmology. The universe
cools as it expands, so it was initially far denser and hotter than it is
today. For hundreds of thousands of years the cosmic gas consisted of a
plasma of free protons and electrons, and a slight mix of light nuclei,
sustained by the intense thermal motion of these particles. Just like the
plasma in our own Sun, the ancient cosmic plasma emitted and scattered a
strong field of visible and ultraviolet photons. As mentioned above, about
$400\,000$ years after the Big Bang the temperature of the universe dipped
for the first time below a few thousand degrees Kelvin. The protons and
electrons were now moving slowly enough that they could attract each other
and form hydrogen atoms, in a process known as cosmic recombination. With
the scattering of the energetic photons now much reduced, the photons
continued traveling in straight lines, mostly undisturbed except that
cosmic expansion has redshifted their wavelength into the microwave regime
today. The emission temperature of the observed spectrum of these CMB
photons is the same in all directions to one part in $100\,000$
(Figure~\ref{fig:CMB}), which reveals that conditions were nearly uniform
in the early universe.

It was just before the moment of cosmic recombination (when matter started
to dominate in energy density over radiation) that gravity started to
amplify the tiny fluctuations in temperature and density observed in the
CMB data. Regions that started out slightly denser than average began to
contract because the gravitational forces were also slightly stronger than
average in these regions. Eventually, after hundreds of millions of years
of contraction, the overdense regions stopped expanding, turned around, and
eventually collapsed to make bound objects such as galaxies. The gas within
these collapsed objects cooled and fragmented into stars. This process,
however, would have taken too long to explain the abundance of galaxies
today, if it involved only the observed cosmic gas. Instead, gravity is
strongly enhanced by the presence of dark matter -- an unknown substance
that makes up the vast majority (83\%) of the cosmic density of matter. The
motion of stars and gas around the centers of nearby galaxies indicates
that each is surrounded by an extended mass of dark matter, and so
dynamically-relaxed dark matter concentrations are generally referred to as
``halos''.

According to the standard cosmological model, the dark matter is cold
(abbreviated as CDM), i.e., it behaves as a collection of collisionless
particles that started out at matter domination with negligible thermal
velocities and have evolved exclusively under gravitational forces. The
model explains how both individual galaxies and the large-scale patterns in
their distribution originated from the small initial density
fluctuations. On the largest scales, observations of the present galaxy
distribution have indeed found the same statistical patterns as seen in the
CMB, enhanced as expected by billions of years of gravitational
evolution. On smaller scales, the model describes how regions that were
denser than average collapsed due to their enhanced gravity and eventually
formed gravitationally-bound halos, first on small spatial scales and later
on larger ones. In this hierarchical model of galaxy formation, the small
galaxies formed first and then merged or accreted gas to form larger
galaxies. At each snapshot of this cosmic evolution, the abundance of
collapsed halos, whose masses are dominated by dark matter, can be computed
from the initial conditions using numerical simulations. The common
understanding of galaxy formation is based on the notion that stars formed
out of the gas that cooled and subsequently condensed to high densities in
the cores of some of these halos.

Gravity thus explains how some gas is pulled into the deep potential wells
within dark matter halos and forms the galaxies. One might naively expect
that the gas outside halos would remain mostly undisturbed. However,
observations show that it has not remained neutral (i.e., in atomic form)
but was largely ionized by the UV radiation emitted by the galaxies. The
diffuse gas pervading the space outside and between galaxies is referred to
as the intergalactic medium (IGM). For the first hundreds of millions of
years after cosmological recombination, the so-called cosmic ``dark ages'',
the universe was filled with diffuse atomic hydrogen.  As soon as galaxies
formed, they started to ionize diffuse hydrogen in their vicinity. Within
less than a billion years, most of the IGM was re-ionized.  We have not yet
imaged the cosmic dark ages before the first galaxies had formed.  One of
the frontiers in current cosmological studies aims to study the cosmic
epoch of reionization and the first generation of galaxies that triggered
it.

\subsection{The expanding universe}

The modern physical description of the Universe as a whole can be traced
back to Einstein, who assumed for simplicity the so-called ``cosmological
principle'': that the distribution of matter and energy is homogeneous and
isotropic on the largest scales. Today isotropy is well established for the
distribution of faint radio sources, optically-selected galaxies, the X-ray
background, and most importantly the cosmic microwave background
(hereafter, CMB). The constraints on homogeneity are less strict, but a
cosmological model in which the Universe is isotropic but significantly
inhomogeneous in spherical shells around our special location, is also
excluded.

In General Relativity, the metric for a space which is spatially
homogeneous and isotropic is the Friedman-Robertson-Walker metric, which
can be written in the form \beq \label{RW}
ds^2=c^2dt^2-a^2(t)\left[\frac{dR^2}{1-k\,R^2}+R^2
\left(d\theta^2+\sin^2\theta\,d\phi^2\right)\right]\ , \eeq where $c$ is
the speed of light, $a(t)$ is the cosmic scale factor which describes
expansion in time $t$, and $(R,\theta,\phi)$ are spherical comoving
coordinates. The constant $k$ determines the geometry of the metric; it is
positive in a closed Universe, zero in a flat Universe, and negative in an
open Universe. Observers at rest remain at rest, at fixed
$(R,\theta,\phi)$, with their physical separation increasing with time in
proportion to $a(t)$. A given observer sees a nearby observer at physical
distance $D$ receding at the Hubble velocity $H(t)D$, where the Hubble
constant at time $t$ is $H(t)=d\,a(t)/dt$. Light emitted by a source at
time $t$ is observed at $t=0$ with a redshift $z=1/a(t)-1$, where we set
$a(t=0) \equiv 1$ for convenience.

The Einstein field equations of General Relativity yield the Friedmann
equation \beq H^2(t)=\frac{8 \pi G}{3}\rho-\frac{k}{a^2}\ , \eeq which
relates the expansion of the Universe to its matter-energy content. The
constant $k$ determines the geometry of the universe; it is positive in a
closed universe, zero in a flat universe, and negative in an open
universe. For each component of the energy density $\rho$, with an equation
of state $p=p(\rho)$, the density $\rho$ varies with $a(t)$ according to
the thermodynamic relation \beq d (\rho c^2 R^3)=-p d(R^3)\ . \eeq With the
critical density \beq \rho_C(t) \equiv \frac{3 H^2(t)}{8 \pi G} \eeq
defined as the density needed for $k=0$, we define the ratio of the total
density to the critical density as \beq \Omega \equiv \frac{\rho}{\rho_C}\
. \eeq With $\Omm$, $\Oml$, and $\Omr$ denoting the present contributions
to $\Omega$ from matter (including cold dark matter as well as a
contribution $\Omega_b$ from ordinary matter [``baryons''] made of protons
and neutrons), vacuum density (cosmological constant), and radiation,
respectively, the Friedmann equation becomes \beq \frac{H(t)}{H_0}= \left[
\frac{\Omm} {a^3}+ \Oml+ \frac{\Omr}{a^4}+ \frac{\Omk}{a^2}\right]\ , \eeq
where we define $H_0$ and $\Omega_0=\Omm+\Oml+\Omr$ to be the present
values of $H$ and $\Omega$, respectively, and we let \beq \Omk \equiv
-\frac{k}{H_0^2}=1-\Omega_m. \eeq In the particularly simple Einstein-de
Sitter model ($\Omm=1$, $\Oml=\Omr=\Omk=0$), the scale factor varies as
$a(t) \propto t^{2/3}$. Even models with non-zero $\Oml$ or $\Omk$ approach
the Einstein-de Sitter scaling-law at high redshift, i.e.\ when $(1+z) \gg
|\Omm^{-1}-1|$ (as long as $\Omr$ can be neglected). In this high-$z$
regime the age of the Universe is
\begin{equation}
t\approx {2\over 3 H_0 \sqrt{\Omega_m}} \left(1+z\right)^{-3/2}\
\approx 10^9 {\rm yr} \left({1+z\over 7}\right)^{-3/2} .
\end{equation}

Recent observations confine the standard set of cosmological parameters to
a relatively narrow range. In particular, we seem to live in a universe
dominated by a cosmological constant ($\Lambda$) and cold dark matter, or
in short a $\Lambda$CDM cosmology (with $\Omk$ so small that it is usually
assumed to equal zero) with an approximately scale-invariant primordial
power spectrum of density fluctuations, i.e., $n \approx 1$ where the
initial power spectrum is $P(k)=\vert \delta_{\bf k} \vert^2\propto k^n$ in
terms of the wavenumber $k$ of the Fourier modes $\delta_{\bf k}$ (see \S
\ref{sec:lin} below). Also, the Hubble constant today is written as
$H_0=100h \mbox{ km s}^{-1}\mbox{Mpc}^{-1}$ in terms of $h$, and the
overall normalization of the power spectrum is specified in terms of
$\sigma_8$, the root-mean-square amplitude of mass fluctuations in spheres
of radius $8\ h^{-1}$ Mpc. For example, the best-fit cosmological
parameters matching the WMAP data together with large-scale gravitational
lensing observations are $\sigma_8=0.826$, $n=0.953$, $h=0.687$,
$\Omega_m=0.299$, $\Omega_\Lambda=0.701$ and $\Omega_b=0.0478$.
%

\section{Galaxy Formation}

\subsection{Growth of linear perturbations}

\label{sec:lin}

As noted in the Introduction, observations of the CMB show that the
universe at cosmic recombination (redshift $z\sim 10^3$) was remarkably
uniform apart from spatial fluctuations in the energy density and in the
gravitational potential of roughly one part in $\sim 10^5$. The primordial
inhomogeneities in the density distribution grew over time and eventually
led to the formation of galaxies as well as galaxy clusters and large-scale
structure. In the early stages of this growth, as long as the density
fluctuations on the relevant scales were much smaller than unity, their
evolution can be understood with a linear perturbation analysis.

As before, we distinguish between fixed and comoving coordinates. Using
vector notation, the fixed coordinate ${\bf r}$ corresponds to a comoving
position $\xb=\rb/a$. In a homogeneous Universe with density $\rho$, we
describe the cosmological expansion in terms of an ideal pressureless fluid
of particles each of which is at fixed $\xb$, expanding with the Hubble
flow $\vb=H(t) \rb$ where $\vb=d\rb/dt$. Onto this uniform expansion we
impose small perturbations, given by a relative density perturbation \beq
\delta(\xb)=\frac{\rho(\rb)}{\bar{\rho}}-1\ , \eeq where the mean fluid
density is $\bar{\rho}$, with a corresponding peculiar velocity $\ub \equiv
\vb - H \rb$. Then the fluid is described by the continuity and Euler
equations in comoving coordinates: \beqa \frac{\partial
\delta}{\partial t}+\frac{1}{a}{\bf \nabla} \cdot \left[(1+\delta)
\ub\right] &=& 0 \\ \frac{\partial \ub}{\partial t}+H\ub+\frac{1}{a}(\ub
\cdot {\bf \nabla}) \ub&=&-\frac{1}{a}{\bf \nabla}\phi\ .  \eeqa The
potential $\phi$ is given by the Poisson equation, in terms of the density
perturbation: \beq \nabla^2\phi=4 \pi G \bar{\rho} a^2 \delta\ . \eeq This
fluid description is valid for describing the evolution of collisionless
cold dark matter particles until different particle streams cross.  This
``shell-crossing'' typically occurs only after perturbations have grown to
become non-linear, and at that point the individual particle trajectories
must in general be followed. Similarly, baryons can be described as a
pressureless fluid as long as their temperature is negligibly small, but
non-linear collapse leads to the formation of shocks in the gas.

For small perturbations $\delta \ll 1$, the fluid equations can be
linearized and combined to yield \beq \frac{\partial^2\delta}{\partial
t^2}+2 H \frac{\partial\delta}{\partial t}=4 \pi G \bar{\rho} \delta\
. \eeq This linear equation has in general two independent solutions, only
one of which grows with time. Starting with random initial conditions, this
``growing mode'' comes to dominate the density evolution. Thus, until it
becomes non-linear, the density perturbation maintains its shape in
comoving coordinates and grows in proportion to a growth factor $D(t)$. The
growth factor in the matter-dominated era is given by \beq D(t)
\propto \frac{\left(\Oml a^3+\Omk a+\Omm\right)^{1/2}}{a^{3/2}}\int_0^a
\frac{a'^{3/2}\, da'}{\left(\Oml a'^3+\Omk a'+\Omm\right)^{3/2}}\ , \eeq
where we neglect $\Omr$ when considering halos forming in the
matter-dominated regime at $z \ll 10^4$. In the Einstein-de Sitter model
(or, at high redshift, in other models as well) the growth factor is simply
proportional to $a(t)$.

The spatial form of the initial density fluctuations can be described in
Fourier space, in terms of Fourier components \beq \delta_\kb = \int d^3x\,
\delta(x) e^{-i \kb \cdot \xb}\ .\eeq Here we use the comoving wave-vector
$\kb$, whose magnitude $k$ is the comoving wavenumber which is equal to
$2\pi$ divided by the wavelength. The Fourier description is particularly
simple for fluctuations generated by inflation. Inflation generates
perturbations given by a Gaussian random field, in which different
$\kb$-modes are statistically independent, each with a random phase. The
statistical properties of the fluctuations are determined by the variance
of the different $\kb$-modes, and the variance is described in terms of the
power spectrum $P(k)$ as follows: \beq \left<\delta_{\kb} \delta_{{\bf
k'}}^{*}\right>=\left(2 \pi\right)^3 P(k)\, \delta^{(3)} \left(\kb-{\bf
k'}\right)\ , \eeq where $\delta^{(3)}$ is the three-dimensional Dirac
delta function.  The gravitational potential fluctuations are sourced by
the density fluctuations through Poisson's equation.

In standard models, inflation produces a primordial power-law spectrum
$P(k) \propto k^n$ with $n \sim 1$. Perturbation growth in the
radiation-dominated and then matter-dominated Universe results in a
modified final power spectrum, characterized by a turnover at a scale
of order the horizon $cH^{-1}$ at matter-radiation equality, and a
small-scale asymptotic shape of $P(k) \propto k^{n-4}$. The overall
amplitude of the power spectrum is not specified by current models of
inflation, and it is usually set by comparing to the observed CMB
temperature fluctuations or to local measures of large-scale
structure.

Since density fluctuations may exist on all scales, in order to determine
the formation of objects of a given size or mass it is useful to consider
the statistical distribution of the smoothed density field.  Using a window
function $W(\rb)$ normalized so that $\int d^3r\, W(\rb)=1$, the smoothed
density perturbation field, $\int d^3r \delta(\xb) W(\rb)$, itself follows
a Gaussian distribution with zero mean. For the particular choice of a
spherical top-hat, in which $W=1$ in a sphere of radius $R$ and is zero
outside, the smoothed perturbation field measures the fluctuations in the
mass in spheres of radius $R$. The normalization of the present power
spectrum is often specified by the value of $\sigma_8 \equiv \sigma(R=8
h^{-1} {\rm Mpc})$. For the top-hat, the smoothed perturbation field is
denoted $\delta_R$ or $\delta_M$, where the mass $M$ is related to the
comoving radius $R$ by $M=4 \pi \rho_m R^3/3$, in terms of the current mean
density of matter $\rho_m$. The variance $\langle \delta_M \rangle^2$ is
\beq \sigma^2(M)= \sigma^2(R)= \int_0^{\infty}\frac{dk}{2 \pi^2} \,k^2 P(k)
\left[\frac{3 j_1(kR)}{kR} \right]^2\ ,\label{eqsigM}\eeq where
$j_1(x)=(\sin x-x \cos x)/x^2$. The function $\sigma(M)$ plays a crucial
role in estimates of the abundance of collapsed objects, as we describe
later.

Different physical processes contributed to the perturbation growth. In the
absence of other influences, gravitational forces due to density
perturbations imprinted by inflation would have driven parallel
perturbation growth in the dark matter, baryons and photons. However, since
the photon sound speed is of order the speed of light, the radiation
pressure produced sound waves on a scale of order the cosmic horizon and
suppressed sub-horizon perturbations in the photon density. The baryonic
pressure similarly suppressed perturbations in the gas below the (much
smaller) so-called baryonic {\it Jeans} scale. Since the formation of
hydrogen at recombination had decoupled the cosmic gas from its mechanical
drag on the CMB, the baryons subsequently began to fall into the
pre-existing gravitational potential wells of the dark matter.

Spatial fluctuations developed in the gas temperature as well as in the gas
density. Both the baryons and the dark matter were affected on small scales
by the temperature fluctuations through the gas pressure.  Compton heating
due to scattering of the residual free electrons (constituting a fraction
$\sim10^{-4}$) with the CMB photons remained effective, keeping the gas
temperature fluctuations tied to the photon temperature fluctuations, even
for a time after recombination.  The growth of linear perturbations can be
calculated with the standard CMBFAST code (http://www.cmbfast.org), after a
modification to account for the fact that the speed of sound of the gas
also fluctuates spatially.

\begin{figure}
\epsfxsize=10cm \epsfbox{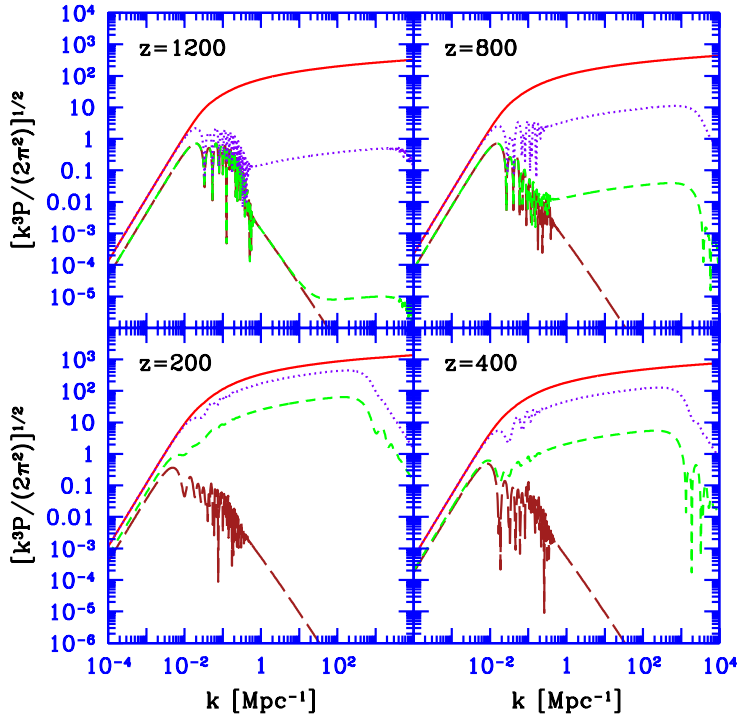}
\caption{Power spectra of density and temperature fluctuations vs.\
comoving wavenumber, at redshifts 1200, 800, 400, and 200 (from Barkana \&
Loeb 2007). We consider fluctuations in the CDM density (solid curves),
baryon density (dotted curves), baryon temperature (short-dashed curves),
and photon temperature (long-dashed curves).}
\label{fig:photons}
\end{figure}

The magnitude of the fluctuations in the CDM and baryon densities, and in
the baryon and photon temperatures, is shown in Figure~\ref{fig:photons},
in terms of the dimensionless combination $[k^3 P(k)/(2 \pi^2)]^{1/2}$,
where $P(k)$ is the corresponding power spectrum of fluctuations in terms
of the comoving wavenumber $k$ of each Fourier mode. After recombination,
two main drivers affect the baryon density and temperature fluctuations,
namely, the thermalization with the CMB and the gravitational force that
attracts the baryons to the dark matter potential wells. As shown in the
figure, the density perturbations in all species grow together on scales
where gravity is unopposed, outside the horizon (i.e., at $k \lesssim 0.01$
Mpc$^{-1}$ at $z \sim 1000$). At $z=1200$ the perturbations in the
baryon-photon fluid oscillate as acoustic waves on scales of order the
sound horizon ($k \sim 0.01~{\rm Mpc^{-1}}$), while smaller-scale
perturbations in both the photons and baryons are damped by photon
diffusion and the drag of the diffusing photons on the baryons. On
sufficiently small scales the power spectra of baryon density and
temperature roughly assume the shape of the dark matter fluctuations
(except for the gas-pressure cutoff at the very smallest scales), due to
the effect of gravitational attraction on the baryon density and of the
resulting adiabatic expansion on the gas temperature. After the mechanical
coupling of the baryons to the photons ends at $z \sim 1000$, the baryon
density perturbations gradually grow towards the dark matter perturbations
because of gravity. Similarly, after the thermal coupling ends at $z \sim
200$, the baryon temperature fluctuations are driven by adiabatic expansion
towards a value of 2/3 of the density fluctuations. As the figure shows, by
$z=200$ the baryon infall into the dark matter potentials is well advanced
and adiabatic expansion is becoming increasingly important in setting the
baryon temperature.

\subsection{Halo properties}

The small density fluctuations evidenced in the CMB grow over time as
described in the previous subsection, until the perturbation $\delta$
becomes of order unity, and the full non-linear gravitational problem
must be considered. The dynamical collapse of a dark matter halo can
be solved analytically only in cases of particular symmetry. If we
consider a region which is much smaller than the horizon $cH^{-1}$,
then the formation of a halo can be formulated as a problem in
Newtonian gravity, in some cases with minor corrections coming from
General Relativity. The simplest case is that of spherical symmetry,
with an initial ($t=t_i\ll t_0$) top-hat of uniform overdensity
$\delta_i$ inside a sphere of radius $R$. Although this model is
restricted in its direct applicability, the results of spherical
collapse have turned out to be surprisingly useful in understanding
the properties and distribution of halos in models based on cold dark
matter.

The collapse of a spherical top-hat perturbation is described by the
Newtonian equation (with a correction for the cosmological constant) \beq
\frac{d^2r}{dt^2}=H_0^2 \Oml\, r-\frac{GM}{r^2}\ , \eeq where $r$ is the
radius in a fixed (not comoving) coordinate frame, $H_0$ is the present-day
Hubble constant, $M$ is the total mass enclosed within radius $r$, and the
initial velocity field is given by the Hubble flow $dr/dt=H(t) r$. The
enclosed $\delta$ grows initially as $\delta_L=\delta_i D(t)/D(t_i)$, in
accordance with linear theory, but eventually $\delta$ grows above
$\delta_L$. If the mass shell at radius $r$ is bound (i.e., if its total
Newtonian energy is negative) then it reaches a radius of maximum expansion
and subsequently collapses. As demonstrated in the previous section, at the
moment when the top-hat collapses to a point, the overdensity predicted by
linear theory is $\delta_L\, = 1.686$ in the Einstein-de Sitter model, with
only a weak dependence on $\Omm$ and $\Oml$. Thus a tophat collapses at
redshift $z$ if its linear overdensity extrapolated to the present day
(also termed the critical density of collapse) is \beq \delta_{\rm
crit}(z)=\frac{1.686}{D(z)}\ ,
\label{deltac} \eeq where we set $D(z=0)=1$.

Even a slight violation of the exact symmetry of the initial perturbation
can prevent the tophat from collapsing to a point. Instead, the halo
reaches a state of virial equilibrium by violent relaxation (phase
mixing). Using the virial theorem $U=-2K$ to relate the potential energy
$U$ to the kinetic energy $K$ in the final state (implying that the virial
radius is half the turnaround radius - where the kinetic energy vanishes),
the final overdensity relative to the critical density at the collapse
redshift is $\Delta_c=18\pi^2 \simeq 178$ in the Einstein-de Sitter model,
modified in a Universe with $\Omm+\Oml=1$ to the fitting formula \beq
\Delta_c=18\pi^2+82 d-39 d^2\ , \eeq where $d\equiv \Ommz-1$ is evaluated
at the collapse redshift, so that \beq \Ommz=\frac{\Omm (1+z)^3}{\Omm
(1+z)^3+\Oml+\Omk (1+z)^2}\ .
\label{Ommz} \eeq

A halo of mass $M$ collapsing at redshift $z$ thus has a virial radius \beq
r_{\rm vir}=0.784 \left(\frac{M}{10^8\ h^{-1} \ M_{\sun} }\right)^{1/3}
\left[\frac{\Omm} {\Ommz}\ \frac{\Delta_c} {18\pi^2}\right]^{-1/3} \left
(\frac{1+z}{10} \right)^{-1}\ h^{-1}\ {\rm kpc}\ , \label{rvir}\eeq and a
corresponding circular velocity, \beq V_c=\left(\frac{G M}{r_{\rm
vir}}\right)^{1/2}= 23.4 \left( \frac{M}{10^8\ h^{-1} \ M_{\sun}
}\right)^{1/3} \left[\frac {\Omm} {\Ommz}\ \frac{\Delta_c}
{18\pi^2}\right]^{1/6} \left( \frac{1+z} {10} \right)^{1/2}\ {\rm km\
s}^{-1}\ . \label{Vceqn} \eeq In these expressions we have assumed a
present Hubble constant written in the form $H_0=100\, h\mbox{ km
s}^{-1}\mbox{Mpc}^{-1}$. We may also define a virial temperature \beq
\label{tvir} T_{\rm vir}=\frac{\mu m_p V_c^2}{2 \kB}=1.98\times 10^4\
\left(\frac{\mu}{0.6}\right) \left(\frac{M}{10^8\ h^{-1} \ M_{\sun}
}\right)^{2/3} \left[ \frac {\Omm} {\Ommz}\ \frac{\Delta_c}
{18\pi^2}\right]^{1/3} \left(\frac{1+z}{10}\right)\ {\rm K} \ , \eeq where
$\mu$ is the mean molecular weight and $m_p$ is the proton mass. Note that
the value of $\mu$ depends on the ionization fraction of the gas; for a
fully ionized primordial gas $\mu=0.59$, while a gas with ionized hydrogen
but only singly-ionized helium has $\mu=0.61$. The binding energy of the
halo is approximately\footnote{The coefficient of $1/2$ in
equation~(\ref{Ebind}) would be exact for a singular isothermal sphere with
$\rho(r)\propto 1/r^2$.} \beq \label{Ebind} E_b= {1\over 2}
\frac{GM^2}{r_{\rm vir}} = 5.45\times 10^{53} \left(\frac{M}{10^8\ h^{-1} \
M_{\sun} }\right)^{5/3} \left[ \frac {\Omm} {\Ommz}\ \frac{\Delta_c}
{18\pi^2}\right]^{1/3} \left(\frac{1+z}{10}\right) h^{-1}\ {\rm erg}\
. \eeq Note that the binding energy of the baryons is smaller by a factor
equal to the baryon fraction $\Omega_b/\Omm$.

Although spherical collapse captures some of the physics governing the
formation of halos, structure formation in cold dark matter models proceeds
hierarchically. At early times, most of the dark matter is in low-mass
halos, and these halos continuously accrete and merge to form high-mass
halos. Numerical simulations of hierarchical halo formation indicate a
roughly universal spherically-averaged density profile for the resulting
halos, though with considerable scatter among different halos. The typical
profile has the form \beq \rho(r)=\frac{3 H_0^2} {8 \pi G} (1+z)^3
\frac{\Omm}{\Ommz} \frac{\delta_c} {\cN x (1+\cN x)^2}\ ,
\label{NFW} \eeq where $x=r/r_{\rm vir}$, and the characteristic density
$\delta_c$ is related to the concentration parameter $\cN$ by \beq
\delta_c=\frac{\Delta_c}{3} \frac{\cN^3} {\ln(1+\cN)-\cN/(1+\cN)} \ . \eeq
The concentration parameter itself depends on the halo mass $M$, at a given
redshift $z$.

\subsection{Formation of the first stars}

Theoretical expectations for the properties of the first galaxies are based
on the standard cosmological model outlined in the Introduction. The
formation of the first bound objects marked the central milestone in the
transition from the initial simplicity (discussed in the previous
subsection) to the present-day complexity. Stars and accreting black holes
output copious radiation and also produced explosions and outflows that
brought into the IGM chemical products from stellar nucleosynthesis and
enhanced magnetic fields. However, the formation of the very first stars,
in a universe that had not yet suffered such feedback, remains a
well-specified problem for theorists.

\begin{figure}
\epsfxsize=7cm \epsfbox{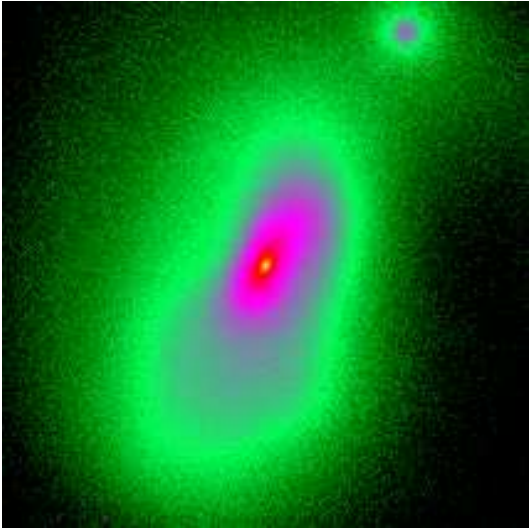}
\caption{Collapse and fragmentation of a primordial cloud of gas (from
Bromm \& Loeb 2004).  Shown is the projected gas density at a redshift
$z\simeq 21.5$, briefly after gravitational runaway collapse has commenced
in the center of the cloud.  
The refined morphology is plotted in a simulation box with linear
physical size of 0.5~pc.  The central density peak, vigorously gaining mass
by accretion, defines the seed of a metal-free (Population III) star.
Searches for metal-poor stars are underway in the halo of the Milky Way
galaxy, an environment less crowded by metal-rich (Population I,II) stars
than the core of our galaxy. The goal of these searches is to constrain
theoretical calculations (such as the one shown here) for the formation of
the first stars.  }
\label{2ab}
\end{figure}

Stars form when huge amounts of matter collapse to enormous
densities. However, the process can be stopped if the pressure exerted by
the hot intergalactic gas prevents outlying gas from falling into dark
matter concentrations. As the gas falls into a dark matter halo, it forms
shocks due to converging supersonic flows and in the process heats up and
can only collapse further by first radiating its energy away. This
restricts this process of collapse to very large clumps of dark matter that
are around $100\,000$ times the mass of the Sun. Inside these clumps, the
shocked gas loses energy by emitting radiation from excited molecular
hydrogen that formed naturally within the primordial gas mixture of
hydrogen and helium.

The first stars are expected to have been quite different from the stars
that form today in the Milky Way. The higher pressure within the primordial
gas due to the presence of fewer cooling agents suggests that fragmentation
only occurred into relatively large units, in which gravity could overcome
the pressure. Due to the lack of carbon, nitrogen, and oxygen -- elements
that would normally dominate the nuclear energy production in modern
massive stars -- the first stars must have condensed to extremely high
densities and temperatures before nuclear reactions were able to heat the
gas and balance gravity. These unusually massive stars produced high
luminosities of UV photons, but their nuclear fuel was exhausted after 2--3
million years, resulting in a huge supernova or in collapse to a black
hole. The heavy elements which were dispersed by the first supernovae in
the surrounding gas, enabled the enriched gas to cool more effectively and
fragment into lower mass stars. Simple calculations indicate that a carbon
or oxygen enrichment of merely $\la 10^{-3}$ of the solar abundance is
sufficient to allow solar mass stars to form. These second-generation
``low-metallicity'' stars are long-lived and could in principle be
discovered in the halo of the Milky Way galaxy, providing fossil record of
the earliest star formation episode in our cosmic environment.

Advances in computing power have made possible detailed numerical
simulations of how the first stars formed. These simulations begin in
the early universe, in which dark matter and gas are distributed uniformly,
apart from tiny variations in density and temperature that are
statistically distributed according to the patterns observed in the CMB. In
order to span the vast range of scales needed to simulate an individual
star within a cosmological context, the latest code 
follows a box 0.3 Mpc in length and zooms in repeatedly on the densest part
of the first collapsing cloud that is found within the simulated
volume. The simulation follows gravity, hydrodynamics, and chemical
processes in the primordial gas, and resolves a scale 10 orders of
magnitudes smaller than that of the simulated box. While the resolved scale
is still three orders of magnitudes larger than the size of the Sun, these
simulations have established that the first stars formed within halos
containing $\sim 10^5 M_{\odot}$ in total mass, and indicate that the first
stars most likely weighed $\sim 100 M_{\odot}$ each.

To estimate {\it when}\/ the first stars formed we must remember that the
first $100\,000$ solar mass halos collapsed in regions that happened to
have a particularly high density enhancement very early on. There was
initially only a small abundance of such regions in the entire universe, so
a simulation that is limited to a small volume is unlikely to find such
halos until much later. Simulating the entire universe is well beyond the
capabilities of current simulations, but analytical models predict that the
first observable star in the universe probably formed 30 million years
after the Big Bang, less than a quarter of one percent of the Universe's
total age of 13.7 billion years.

Although stars were extremely rare at first, gravitational collapse
increased the abundance of galactic halos and star formation sites with
time (Figure~\ref{fig:history}). Radiation from the first stars is expected
to have eventually dissociated all the molecular hydrogen in the
intergalactic medium, leading to the domination of a second generation of
larger galaxies where the gas cooled via radiative transitions in atomic
hydrogen and helium. Atomic cooling occurred in halos of mass above
$\sim10^8 M_{\odot}$, in which the infalling gas was heated above 10,000 K
and became ionized. The first galaxies to form through atomic cooling are
expected to have formed around redshift 45, and such
galaxies were likely the main sites of star formation by the time
reionization began in earnest. As the IGM was heated above 10,000 K by
reionization, its pressure jumped and prevented the gas from accreting into
newly forming halos below $\sim10^9 M_{\odot}$. The first Milky-Way-sized
halo $M = 10^{12} M_{\odot}$ is predicted to have formed 400 million years
after the Big Bang, but such halos have become typical
galactic hosts only in the last five billion years.

Hydrogen is the most abundant element in the Universe, The prominent
Lyman-$\alpha$ spectral line of hydrogen (corresponding to a transition
from its first excited level to its ground state) provides an important
probe of the condensation of primordial gas into the first galaxies.
Existing searches for Lyman-$\alpha$ emission have discovered galaxies
robustly out to a redshift $z\sim 7$ with some unconfirmed candidate
galaxies out to $z\sim 10$. The spectral break owing to Lyman-$\alpha$
absorption by the IGM allows to identify high-redshifts galaxies
photometrically.  Existing observations provide only a preliminary glimpse
into the formation of the first galaxies.

\begin{figure}
\epsfxsize=10cm \epsfbox{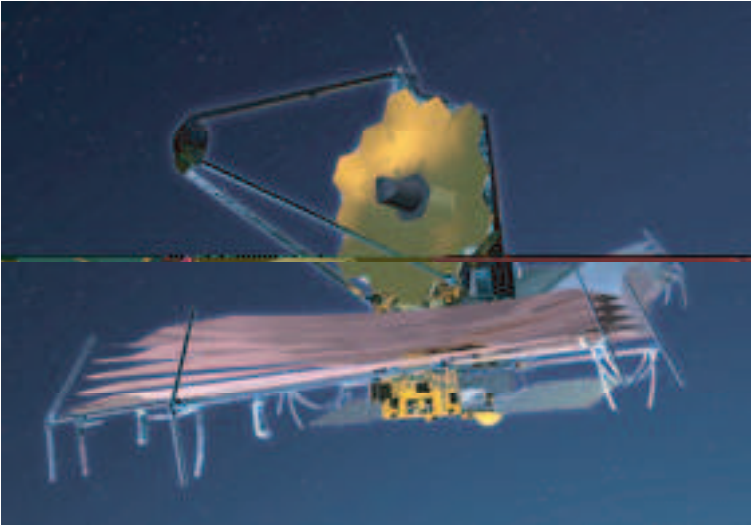}
\caption{A sketch of the current design for the {\it James Webb Space
Telescope}, the successor to the {\it Hubble Space Telescope}\/ to be
launched in 2013 (see http://www.jwst.nasa.gov/). The current design
includes a primary mirror made of beryllium which is 6.5 meters in
diameter as well as an instrument sensitivity that spans the full
range of infrared wavelengths of 0.6--28$\mu$m and will allow
detection of some of the first galaxies in the infant Universe. The
telescope will orbit 1.5 million km from Earth at the Lagrange L2
point. Note that the sun shield (the large flat screen in the image)
is 22m$\times$10m in size.}
\label{jwst}
\end{figure}

Within the next decade, NASA plans to launch an infrared space telescope
({\it JWST}\/; Figure~\ref{jwst}) that will image some of the earliest
sources of light (stars and black holes) in the Universe. In parallel,
there are several initiatives to construct large-aperture infrared
telescopes on the ground with the same goal in mind (see
http://www.eso.org/public/astronomy/projects/e-elt.html;
http://www.tmt.org/;
http://www.gmto.org/).

\begin{figure}
\epsfxsize=10cm \epsfbox{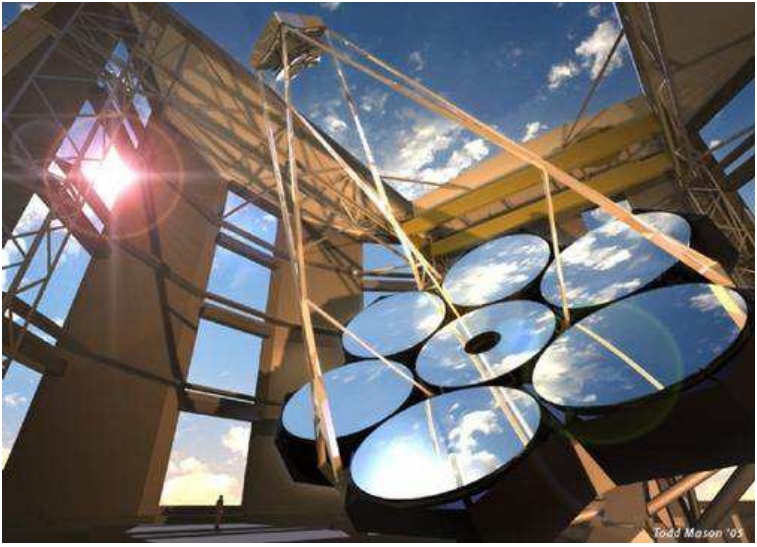}
\caption{Artist's conception of the design for one of the future giant
telescopes that might detect the first generation of galaxies from the
ground. The {\it Giant Magellan Telescope}\/ ({\it GMT}\/) is designed
to contain seven mirrors (each 8.4 meter in diameter) and to have a
resolving power equivalent to a 24.5 meter (80 foot) primary
mirror. For more details see http://www.gmto.org/ .
Two other teams are designing 
competing large telescopes, namely the {\it Thirty Meter Telescope}
(see http://www.tmt.org/)
and the {\it European Extremely Large Telescope} (see
http://www.eso.org/public/astronomy/projects/e-elt.html).
}
\label{gmt}
\end{figure}

The next generation of ground-based telescopes will have a diameter of
twenty to thirty meters (Figure~\ref{gmt}). Together with {\it JWST}\/
(which will not be affected by the atmospheric background) they will be
able to image and make spectral studies of the early galaxies. Given that
these galaxies also create the ionized bubbles around them by their UV
emission, during reionization the locations of galaxies should correlate
with bubbles within the neutral hydrogen.  Within a decade it should be
possible to explore the environmental influence of individual galaxies by
using these telescopes in combination with 21-cm probes of reionization.

\subsection{Gamma-ray Bursts: probing the first stars one 
star at a time}

So far, to learn about diffuse IGM gas pervading the space outside and
between galaxies, astronomers routinely study its absorption signatures in
the spectra of distant quasars, the brightest long-lived astronomical
objects. Quasars' great luminosities are believed to be powered by
accretion of gas onto black holes weighing up to a few billion times the
mass of the Sun that are situated in the centers of massive galaxies. As
the surrounding gas spirals in toward the black hole sink, the viscous
dissipation of heat makes the gas glow brightly into space, creating a
luminous source visible from afar.

\begin{figure}
\epsfxsize=10cm \epsfbox{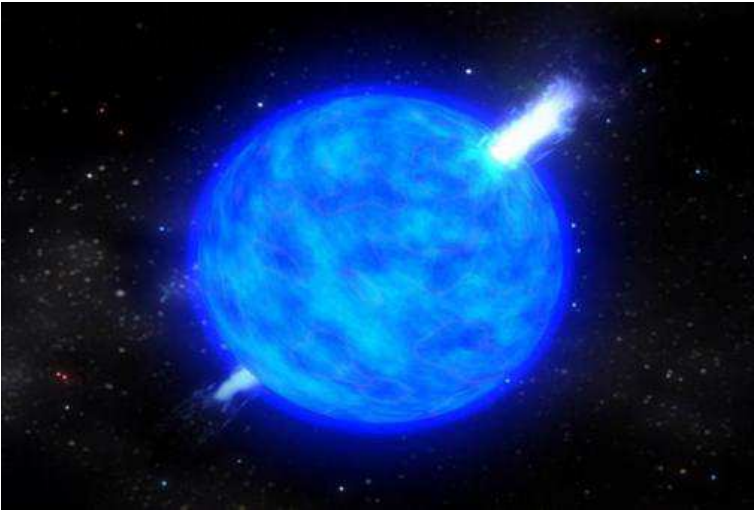}
\caption{Illustration of a long-duration gamma-ray burst in the popular
``collapsar'' model. 
The collapse of the core of a massive star (which lost its hydrogen
envelope) to a black hole generates two opposite jets moving out at a speed
close to the speed of light. The jets drill a hole in the star and shine
brightly towards an observer who happens to be located within with the
collimation cones of the jets. The jets emanating from a single massive
star are so bright that they can be seen across the Universe out to the
epoch when the first stars formed (for more information see
http://swift.gsfc.nasa.gov/). }
\label{grb}
\end{figure}

Over the past decade, an alternative population of bright sources at
cosmological distances was discovered, the so-called afterglows of {\it
Gamma-Ray Bursts} (GRBs). These events are characterized by a flash of
high-energy ($>0.1$ MeV) photons, typically lasting 0.1--100 seconds, which
is followed by an afterglow of lower-energy photons over much longer
timescales. The afterglow peaks at X-ray, UV, optical and eventually radio
wavelengths on time scales of minutes, hours, days, and months,
respectively.  The central engines of GRBs are believed to be associated
with the compact remnants (neutron stars or stellar-mass black holes) of
massive stars. Their high luminosities make them detectable out to the edge
of the visible Universe. GRBs offer the opportunity to detect the most
distant (and hence earliest) population of massive stars, the so-called
Population~III (or Pop~III), one star at a time (Figure~\ref{grb}). In the
hierarchical assembly process of halos that are dominated by cold dark
matter (CDM), the first galaxies should have had lower masses (and lower
stellar luminosities) than their more recent counterparts. Consequently,
the characteristic luminosity of galaxies or quasars is expected to decline
with increasing redshift. GRB afterglows, which already produce a peak flux
comparable to that of quasars or starburst galaxies at $z\sim 1-2$, are
therefore expected to outshine any competing source at the highest
redshifts, when the first dwarf galaxies formed in the Universe.

GRBs, the electromagnetically-brightest explosions in the Universe, should
be detectable out to redshifts $z>10$. High-redshift GRBs can be identified
through infrared photometry, based on the Lyman-$\alpha$ break induced by
absorption of their spectrum at wavelengths below $1.216\, \mu {\rm m}\,
[(1+z)/10]$. Follow-up spectroscopy of high-redshift candidates can then be
performed on a 10-meter-class telescope. GRB afterglows offer the
opportunity to detect stars as well as to probe the metal enrichment level
of the intervening IGM. Recently, the {\it Swift} satellite has
detected a GRB originating at $z\simeq 6.3$, thus demonstrating the
viability of GRBs as probes of the early Universe.

\begin{figure}
\epsfxsize=8cm \epsfbox{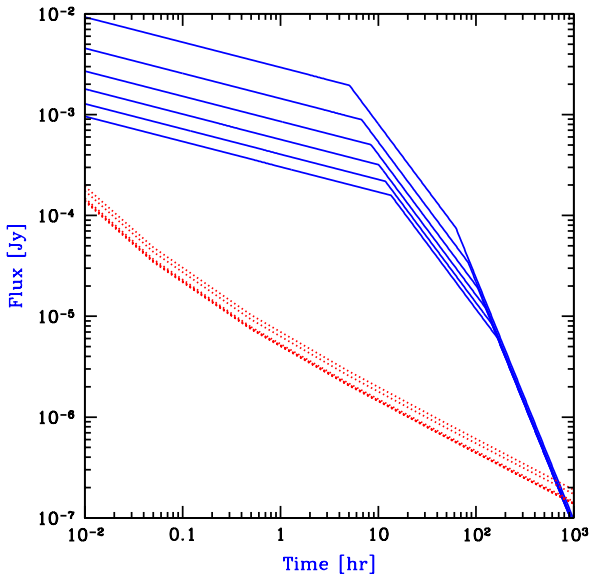}
\caption{GRB afterglow flux as a function of time since the $\gamma$-ray
trigger in the observer frame (from Barkana \& Loeb 2004a). The flux
(solid curves) is calculated at the redshifted Lyman-$\alpha$
wavelength. The dotted curves show the planned detection threshold for the
{\it James Webb Space Telescope} ({\it JWST}), assuming a spectral
resolution $R=5000$ with the near infrared spectrometer, a signal to noise
ratio of 5 per spectral resolution element, and an exposure time equal to
$20\%$ of the time since the GRB explosion (see
http://www.ngst.stsci.edu/nms/main/~).  Each set of curves shows a sequence
of redshifts, namely $z=5$, 7, 9, 11, 13, and 15, respectively, from top to
bottom.}
\label{fig:GRB}
\end{figure}

Another advantage of GRBs is that the GRB afterglow flux at a given
observed time lag after the $\gamma$-ray trigger is not expected to fade
significantly with increasing redshift, since higher redshifts translate to
earlier times in the source frame, during which the afterglow is
intrinsically brighter. For standard afterglow
lightcurves and spectra, the increase in the luminosity distance with
redshift is compensated by this {\it cosmological time-stretching} effect
as shown in Figure~\ref{fig:GRB}.

GRB afterglows have smooth (broken power-law) continuum spectra unlike
quasars which show strong spectral features (such as broad emission lines
or the so-called ``blue bump'') that complicate the extraction of IGM
absorption features. In particular, the extrapolation into the spectral
regime marked by the IGM Lyman-$\alpha$ absorption during the epoch of
reionization is much more straightforward for the smooth UV spectra of GRB
afterglows than for quasars with an underlying broad Lyman-$\alpha$
emission line. However, the interpretation may be complicated by the
presence of damped Lyman-$\alpha$ absorption by dense neutral hydrogen in
the immediate environment of the GRB within its host galaxy. Since GRBs
originate from the dense environment of active star formation, such damped
absorption is expected and indeed has been seen, including in the most
distant GRB at $z=6.3$.

\subsection{Supermassive black holes}

The fossil record in the present-day Universe indicates that every bulged
galaxy hosts a supermassive black hole (BH) at its center. This conclusion
is derived from a variety of techniques which probe the dynamics of stars
and gas in galactic nuclei.  The inferred BHs are dormant or faint most of
the time, but occasionally flash in a short burst of radiation that lasts
for a small fraction of the age of the Universe. The short duty cycle
accounts for the fact that bright quasars are much less abundant than their
host galaxies, but it begs the more fundamental question: {\it why is the
quasar activity so brief?}  A natural explanation is that quasars are
suicidal, namely the energy output from the BHs regulates their own growth.

Supermassive BHs make up a small fraction, $< 10^{-3}$, of the total mass
in their host galaxies, and so their direct dynamical impact is limited to
the central star distribution where their gravitational influence
dominates. Dynamical friction on the background stars keeps the BH close to
the center. Random fluctuations in the distribution of stars induces a
Brownian motion of the BH. This motion can be described by the same Langevin
equation that captures the motion of a massive dust particle as it responds
to random kicks from the much lighter molecules of air around it.  The
characteristic speed by which the BH wanders around the center is small,
$\sim (m_\star/M_{\rm BH})^{1/2}\sigma_\star$, where $m_\star$ and $M_{\rm
BH}$ are the masses of a single star and the BH, respectively, and
$\sigma_\star$ is the stellar velocity dispersion. Since the random force
fluctuates on a dynamical time, the BH wanders across a region that is
smaller by a factor of $\sim (m_\star/M_{\rm BH})^{1/2}$ than the region
traversed by the stars inducing the fluctuating force on it.

The dynamical insignificance of the BH on the global galactic scale is
misleading. The gravitational binding energy per rest-mass energy of
galaxies is of order $\sim (\sigma_\star/c)^2< 10^{-6}$.  Since BH are
relativistic objects, the gravitational binding energy of material that
feeds them amounts to a substantial fraction its rest mass energy. Even if
the BH mass amounts to a fraction as small as $\sim 10^{-4}$ of the
baryonic mass in a galaxy, and only a percent of the accreted rest-mass
energy is deposited into the gaseous environment of the BH, this slight
deposition can unbind the entire gas reservoir of the host galaxy. This
order-of-magnitude estimate explains why quasars may be short lived.  As
soon as the central BH accretes large quantities of gas so as to
significantly increase its mass, it releases large amounts of energy and
momentum that could suppress further accretion onto it. In short, the BH
growth might be {\it self-regulated}.

\begin{figure}
\epsfxsize=12.5cm \epsfbox{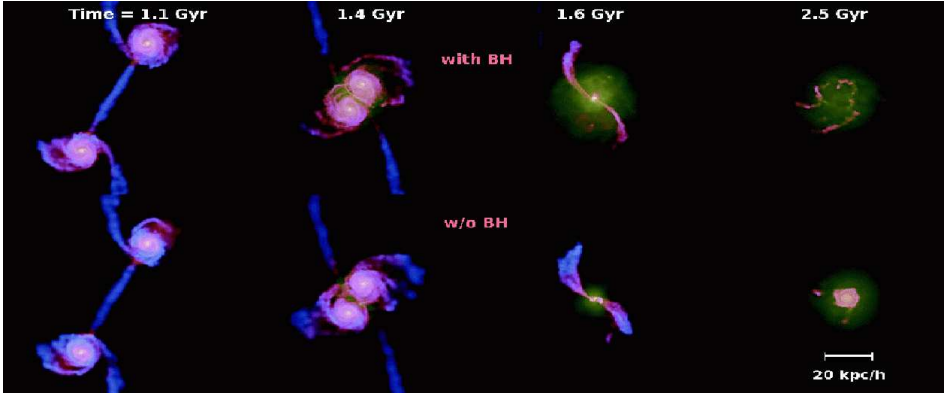}
\caption{Simulation images of a merger of galaxies resulting in quasar
activity that eventually shuts-off the accretion of gas onto the black hole
(from Di Matteo et al. 2005). The upper (lower) panels show a sequence of
snapshots of the gas distribution during a merger with (without) feedback
from a central black hole. The temperature of the gas is color coded.}
\label{merger}
\end{figure}

The principle of {\it self-regulation} naturally leads to a correlation
between the final BH mass, $M_{\rm bh}$, and the depth of the gravitational
potential well to which the surrounding gas is confined. The latter can be
characterized by the velocity dispersion of the associated stars, $\sim
\sigma_\star^2$. Indeed a correlation between $M_{\rm bh}$ and
$\sigma_\star^4$ is observed in the present-day Universe.
If quasars shine near
their Eddington limit as suggested by observations of low and high-redshift
quasars, then 
a fraction of $\sim 5$--$10\%$ of the energy released by the quasar over a
galactic dynamical time needs to be captured in the surrounding galactic
gas in order for the BH growth to be self-regulated (see Figure
\ref{merger}).  With this interpretation, the $M_{\rm bh}$--$\sigma_\star$
relation reflects the limit introduced to the BH mass by self-regulation;
deviations from this relation are inevitable during episodes of BH growth
or as a result of mergers of galaxies that have no cold gas in them.  A
physical scatter around this upper envelope could also result from
variations in the efficiency by which the released BH energy couples to the
surrounding gas.

Various prescriptions for self-regulation were sketched in the
literature. These involve either energy or momentum-driven winds, with the
latter type being a factor of $\sim v_c/c$ ~less efficient.  The quasar
remains active during the dynamical time of the initial gas reservoir,
$\sim 10^7$ years, and fades afterwards due to the dilution of this
reservoir.  The BH growth may resume if the cold gas reservoir is
replenished through a new merger.  Following early analytic work, extensive
numerical simulations demonstrated that galaxy mergers do produce the
observed correlations between black hole mass and spheroid
properties. Because of the limited resolution near the galaxy nucleus,
these simulations adopt a simple prescription for the accretion flow that
feeds the black hole. The actual feedback in reality may depend crucially
on the geometry of this flow and the physical mechanism that couples the
energy or momentum output of the quasar to the surrounding gas.

The inflow of cold gas towards galaxy centers during the growth phase of
the BH would naturally be accompanied by a burst of star formation.  The
fraction of gas that is not consumed by stars or ejected by
supernova-driven winds, will continue to feed the BH. It is therefore not
surprising that quasar and starburst activities co-exist in Ultra Luminous
Infrared Galaxies, and that all quasars show broad metal lines indicating 
pre-enrichment of the surrounding gas with heavy elements.

The upper mass of galaxies may also be regulated by the energy output from
quasar activity. This would account for the fact that cooling flows are
suppressed in present-day X-ray clusters, and that massive BHs and stars in
galactic bulges were already formed at $z\sim 2$. 
In the cores of cooling X-ray clusters, there is often an active central BH
that supplies sufficient energy to compensate for the cooling of the
gas. The primary physical process by which this energy couples to the gas
is still unknown.

The quasars discovered so far at $z\sim 6$ mark the early growth of the
most massive BHs and galactic spheroids. The BHs powering these bright
quasars possess a mass of a few billion solar masses.  A quasar radiating
at its Eddington limiting luminosity, $L_E=1.4\times 10^{47}~{\rm
erg~s^{-1}}(M_{\rm bh}/10^9M_\odot)$, with a radiative efficiency,
$\epsilon_{\rm rad}=L_{E}/{\dot M}c^2$, for converting accreted mass into
radiation, would grow exponentially in mass as a function of time $t$,
$M_{\rm bh} =M_{\rm seed}\exp\{t/t_E\}$ from its initial seed mass $M_{\rm
seed}$, on a time scale, $t_E=4.1\times 10^7~{\rm yr} (\epsilon_{\rm
rad}/0.1)$. Thus, the required growth time in units of the Hubble time
$t_{\rm hubble}= 10^9~{\rm yr}[(1+z)/7]^{-3/2}$ is
\begin{equation}
{t_{\rm growth}\over t_{\rm hubble}}=0.7 \left({\epsilon_{\rm rad} \over
10\%}\right) \left({1+z\over 7}\right)^{3/2}\ln \left({{M_{\rm
bh}/10^9M_\odot} \over M_{\rm seed}/100M_\odot}\right) ~.
\end{equation}
The age of the Universe at $z\sim 6$ provides just sufficient time to grow
a BH with $M_{\rm bh}\sim 10^9M_\odot$ out of a stellar mass seed with
$\epsilon_{\rm rad}=10 \%$. The growth time is shorter for smaller
radiative efficiencies or a higher seed mass.

\subsection{The epoch of reionization}

\begin{figure}
\epsfxsize=8cm \epsfbox{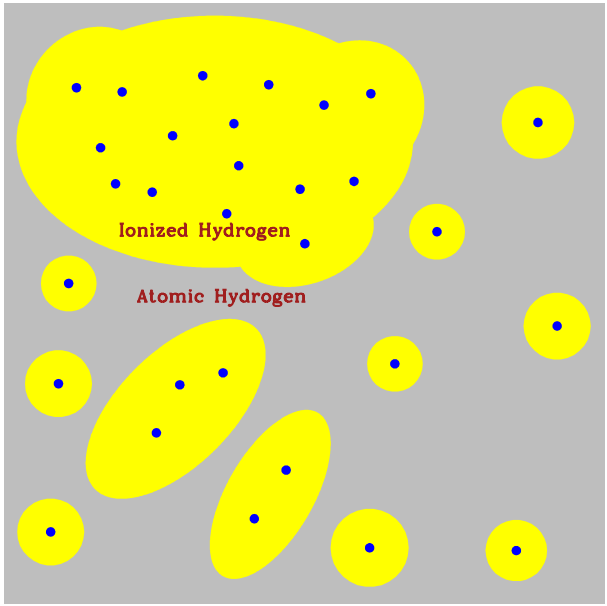}
\caption{The spatial structure of cosmic reionization. The illustration
shows how regions with large-scale overdensities form large concentrations
of galaxies (dots) whose ionizing photons produce enormous joint ionized
bubbles (upper left). At the same time, galaxies are rare within
large-scale voids, in which the IGM is still mostly neutral (lower
right). 
}
\label{fig:rei}
\end{figure}

Given the understanding described above of how many galaxies formed at
various times, the course of reionization can be determined universe-wide
by counting photons from all sources of light. Both stars and black holes
contribute ionizing photons, but the early universe is dominated by small
galaxies which in the local universe have central black holes that are
disproportionately small, and indeed quasars are rare above redshift
6. Thus, stars most likely dominated the production of ionizing UV photons
during the reionization epoch [although high-redshift galaxies should have
also emitted X-rays from accreting black holes and accelerated particles in
collisionless shocks]. Since most stellar ionizing photons are only
slightly more energetic than the 13.6 eV ionization threshold of hydrogen,
they are absorbed efficiently once they reach a region with substantial
neutral hydrogen). This makes the IGM during reionization a two-phase
medium characterized by highly ionized regions separated from neutral
regions by sharp ionization fronts (see Figure~\ref{fig:rei}).

We can obtain a first estimate of the requirements of reionization by
demanding one stellar ionizing photon for each hydrogen atom in the
IGM. If we conservatively assume that stars within the early galaxies
were similar to those observed locally, then each star produced $\sim
4000$ ionizing photons per baryon. Star formation is observed today to
be an inefficient process, but even if stars in galaxies formed out of
only $\sim10\%$ of the available gas, it was still sufficient to
accumulate a small fraction (of order $0.1\%$) of the total baryonic
mass in the universe into galaxies in order to ionize the entire
IGM. More accurate estimates of the actual required fraction account
for the formation of some primordial stars (which were massive,
efficient ionizers, as discussed above), and for recombinations of
hydrogen atoms at high redshifts and in dense regions.

From studies of quasar absorption lines at $z\sim 6$ we know that the IGM
is highly ionized a billion years after the big bang. There are hints,
however, that some large neutral hydrogen regions persist at these early
times and so this suggests that we may not need to go to much higher
redshifts to begin to see the epoch of reionization.  We now know that the
universe could not have fully reionized earlier than an age of 300 million
years, since WMAP observed the effect of the freshly created plasma at
reionization on the large-scale polarization anisotropies of the CMB and
this limits the reionization redshift; an earlier reionization, when the
universe was denser, would have created a stronger scattering signature
that would be inconsistent with the WMAP observations. In any case, the
redshift at which reionization ended only constrains the overall cosmic
efficiency of ionizing photon production. In comparison, a detailed picture
of reionization as it happens will teach us a great deal about the
population of young galaxies that produced this cosmic phase transition.
A key point is that the spatial distribution of ionized bubbles is
determined by clustered groups of galaxies and not by individual
galaxies. At such early times galaxies were strongly clustered even on very
large scales (up to tens of Mpc), and these scales therefore dominate the
structure of reionization. The basic idea is simple. At high redshift,
galactic halos are rare and correspond to rare, high density peaks. As an
analogy, imagine searching on Earth for mountain peaks above 5000
meters. The 200 such peaks are not at all distributed uniformly but instead
are found in a few distinct clusters on top of large mountain ranges. Given
the large-scale boost provided by a mountain range, a small-scale crest
need only provide a small additional rise in order to become a 5000 meter
peak. The same crest, if it formed within a valley, would not come anywhere
near 5000 meters in total height. Similarly, in order to find the early
galaxies, one must first locate a region with a large-scale density
enhancement, and then galaxies will be found there in abundance.

The ionizing radiation emitted from the stars in each galaxy initially
produces an isolated ionized bubble. However, in a region dense with
galaxies the bubbles quickly overlap into one large bubble, completing
reionization in this region while the rest of the universe is still mostly
neutral (Figure~\ref{fig:rei}). Most importantly, since the abundance of
rare density peaks is very sensitive to small changes in the density
threshold, even a large-scale region with a small enhanced density (say,
10\% above the mean density of the universe) can have a much larger
concentration of galaxies than in other regions (e.g., a 50\%
enhancement). On the other hand, reionization is harder to achieve in dense
regions, since the protons and electrons collide and recombine more often
in such regions, and newly-formed hydrogen atoms need to be reionized again
by additional ionizing photons. However, the overdense regions end up
reionizing first since the number of ionizing sources in these regions is
increased so strongly. The large-scale topology of reionization is
therefore inside out, with underdense voids reionizing only at the very end
of reionization, with the help of extra ionizing photons coming in from
their surroundings (which have a higher density of galaxies than the voids
themselves). This is a key prediction awaiting observational testing.

Detailed analytical models that account for large-scale variations in the
abundance of galaxies confirm that the typical bubble size starts well
below a Mpc early in reionization, as expected for an individual galaxy,
rises to 5--10 Mpc during the central phase (i.e., when the universe is
half ionized), and then by another factor of $\sim$5 towards the end of
reionization. These scales are given in comoving units that scale with the
expansion of the universe, so that the actual sizes at a redshift $z$ were
smaller than these numbers by a factor of $1+z$. Numerical simulations have
only recently begun to reach the enormous scales needed to capture this
evolution. Accounting precisely for gravitational evolution on a wide range
of scales but still crudely for gas dynamics, star formation, and the
radiative transfer of ionizing photons, the simulations confirm that the
large-scale topology of reionization is inside out, and that this topology
can be used to study the abundance and clustering of the ionizing sources
(Figures~\ref{fig:rei}).

The characteristic observable size of the ionized bubbles at the end
reionization can be calculated based on simple considerations that only
depend on the power-spectrum of density fluctuations and the redshift. As
the size of an ionized bubble increases, the time it takes a 21-cm photon
emitted by hydrogen to traverse it gets longer. At the same time, the
variation in the time at which different regions reionize becomes smaller
as the regions grow larger. Thus, there is a maximum size above which the
photon crossing time is longer than the cosmic variance in ionization
time. Regions bigger than this size will be ionized at their near side by
the time a 21-cm photon will cross them towards the observer from their far
side. They would appear to the observer as one-sided, and hence signal the
end of reionization. These considerations imply a characteristic size for
the ionized bubbles of $\sim 10$ physical Mpc at $z\sim 6$ (equivalent to
70 Mpc today).  This result implies that future radio experiments should be
tuned to a characteristic angular scale of tens of arcminutes for an
optimal detection of 21-cm brightness fluctuations near the end of
reionization (see \S \ref{sec:21-cmAtomic}).

\subsection{Post-reionization suppression of low-mass galaxies}

After the ionized bubbles overlapped in each region, the ionizing
background increased sharply, and the IGM was heated by the ionizing
radiation to a temperature $T_{\rm IGM}\gtrsim 10^4$ K. Due to the
substantial increase in the IGM pressure, the smallest mass scale into
which the cosmic gas could fragment, the so-called Jeans mass,
increased dramatically, changing the minimum mass of forming galaxies.

Gas infall depends sensitively on the Jeans mass. When a halo more massive
than the Jeans mass begins to form, the gravity of its dark matter
overcomes the gas pressure. Even in halos below the Jeans mass, although
the gas is initially held up by pressure, once the dark matter collapses
its increased gravity pulls in some gas. Thus, the Jeans mass is generally
higher than the actual limiting mass for accretion. Before reionization,
the IGM is cold and neutral, and the Jeans mass plays a secondary role in
limiting galaxy formation compared to cooling. After reionization, the
Jeans mass is increased by several orders of magnitude due to the
photoionization heating of the IGM, and hence begins to play a dominant
role in limiting the formation of stars.  Gas infall in a reionized and
heated Universe has been investigated in a number of numerical simulations.
Three dimensional numerical simulations found a significant suppression of
gas infall in even larger halos ($V_c \sim 75\ {\rm km\ s}^{-1}$), but this
was mostly due to a suppression of late infall at $z\la 2$.

When a volume of the IGM is ionized by stars, the gas is heated to a
temperature $T_{\rm IGM}\sim 10^4$ K. If quasars dominate the UV background
at reionization, their harder photon spectrum leads to $T_{\rm IGM}>
2\times 10^4$ K. Including the effects of dark matter, a given temperature
results in a linear Jeans mass corresponding to a halo circular velocity of
\beq V_J\approx 80 \left(\frac{T_{\rm IGM}}{1.5\times 10^4 {\rm
K}}\right)^{1/2}\ {\rm km\ s}^{-1}. \eeq In halos with a circular
velocity well above $V_J$, the gas fraction in infalling gas equals
the universal mean of $\Omega_b/\Omega_m$, but gas infall is
suppressed in smaller halos.  A simple estimate of the limiting
circular velocity, below which halos have essentially no gas infall,
is obtained by substituting the virial overdensity for the mean
density in the definition of the Jeans mass. The resulting estimate is
\beq V_{\rm lim}=34 \left(\frac{T_{\rm IGM}}{1.5\times 10^4 {\rm
K}}\right)^{1/2}\ {\rm km\ s}^{-1}. \eeq This value is in rough
agreement with the numerical simulations mentioned before.  

Although the Jeans mass is closely related to the rate of gas infall at a
given time, it does not directly yield the total gas residing in halos at a
given time. The latter quantity depends on the entire history of gas
accretion onto halos, as well as on the merger histories of halos, and an
accurate description must involve a time-averaged Jeans mass. 
The gas content of halos in simulations is well fit by an expression which
depends on the filtering mass, a particular time-averaged Jeans mass.

The reionization process was not perfectly synchronized throughout the
Universe. Large-scale regions with a higher density than the mean tended to
form galaxies first and reionized earlier than underdense regions. The
suppression of low-mass galaxies by reionization is therefore modulated by
the fluctuations in the timing of reionization.  Inhomogeneous reionization
imprint a signature on the power-spectrum of low-mass galaxies. Future
high-redshift galaxy surveys hoping to constrain inflationary parameters
must properly model the effects of reionization; conversely, they will also
place new constraints on the thermal history of the IGM during
reionization.

\section{Probing the Diffuse Intergalactic Hydrogen}

\label{sec:absorb}

\subsection{Lyman-alpha absorption}

Resonant Lyman-$\alpha$ absorption has thus far proved to be the best probe
of the state of the IGM. The optical depth to absorption by a uniform
intergalactic medium is \beqa \tau_{s}&=&{\pi e^2 f_\alpha \lambda_\alpha
n_{\HI}(z) \over m_e cH(z)} \label{G-P} \\ \nonumber &\approx& 6.45\times
10^5 x_{\HI} \left({\Omega_bh\over 0.0315}\right)\left({\Omega_m\over
0.3}\right)^{-1/2} \left({1+z\over 10}\right)^{3/2}\ , \eeqa where
$H\approx 100h~{\rm km~s^{-1}~Mpc^{-1}} \Omega_m^{1/2} (1+z)^{3/2}$ is the
Hubble parameter at redshift $z$; $f_\alpha=0.4162$ and
$\lambda_\alpha=1216$\AA\, are the oscillator strength and the wavelength
of the Lyman-$\alpha$ transition; $n_{\HI}(z)$ is the neutral hydrogen
density at $z$ (assuming primordial abundances); $\Omega_m$ and $\Omega_b$
are the present-day density parameters of all matter and of baryons,
respectively; and $x_{\HI}$ is the average fraction of neutral hydrogen. In
the second equality we have implicitly considered high redshifts.

\begin{figure}
\epsfxsize=12cm \epsfbox{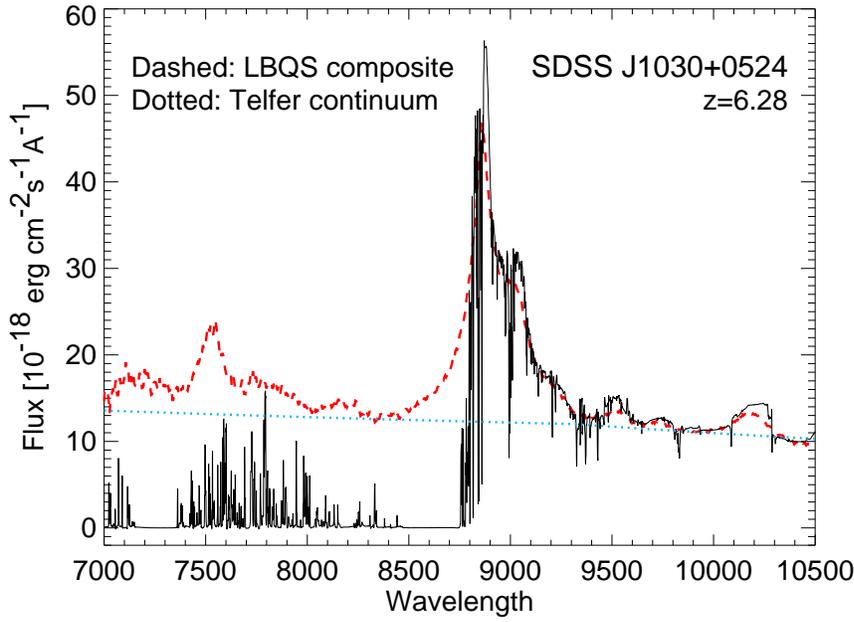}
\caption{Using Lyman-$\alpha$ absorption in quasar spectra to probe the
ionization state of the IGM. This figure from White et al. (2003) shows the
observed spectrum of a $z = 6.28$ quasar (solid curve), and the expected
unabsorbed emission (dashed curve), based on an average over many quasars
seen at lower redshifts. The unabsorbed emission is a sum of smooth
emission (the ``continuum'', dotted curve) plus emission features from
atomic resonances (``emission lines'').  }
\label{fig:white}
\end{figure}

Lyman-$\alpha$ absorption is thus highly sensitive to the presence of even
trace amounts of neutral hydrogen. The lack of full absorption in quasar
spectra then implies that the IGM has been very highly ionized during much
of the history of the universe, from at most a billion years after the big
bang to the present time. At redshifts approaching six, however, the
optical depth increases, and the observed absorption becomes very strong.
An example of this is shown in Figure~\ref{fig:white}, where an observed
quasar spectrum is compared to the unabsorbed expectation for the same
quasar. The prominent Lyman-$\alpha$ emission line, which is produced by
radiating hot gas near the quasar itself, is centered at a wavelength of
8850\AA, which for the redshift (6.28) of this quasar corresponds to a
rest-frame 1216\AA. Above this wavelength, the original emitted quasar
spectrum is seen, since photons emitted with wavelengths higher than
1216\AA\ redshift to higher wavelengths during their journey toward us and
never encounter resonance lines of hydrogen atoms. Shorter-wavelength
photons, however, redshift until they hit the local 1216\AA\ and are then
absorbed by any existing hydrogen atoms. The difference between the
unabsorbed expectation and the actual observed spectrum can be used to
measure the amount of absorption, and thus to infer the atomic hydrogen
density. For this particular quasar, this difference is very large (i.e.,
the observed flux is near zero) just to the blue of the Lyman-$\alpha$
emission line.

\begin{figure}
\epsfxsize=10cm \epsfbox{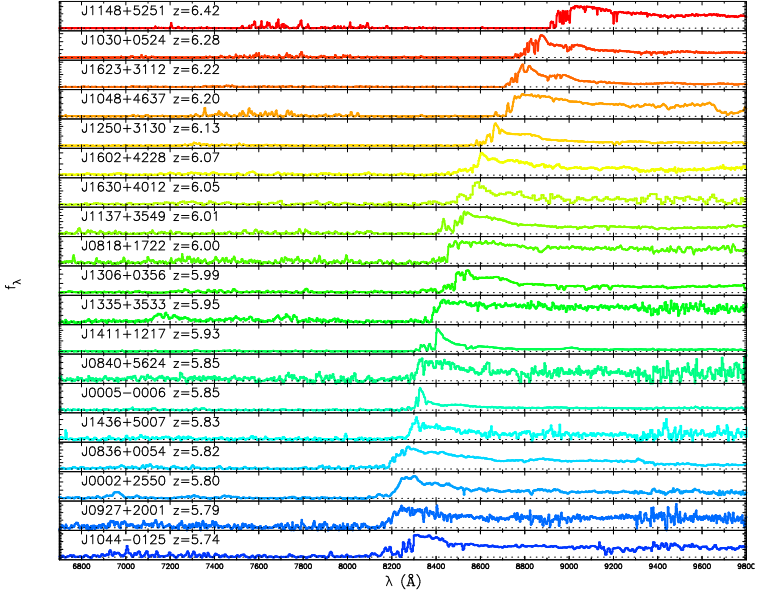}
\caption{Spectra of 19 quasars with redshifts $5.74<z<6.42$ from the {\it
Sloan Digital Sky Survey}, taken from Fan et al. (2005). For some of the
highest-redshift quasars, the spectrum shows no transmitted flux shortward
of the Lyman-$\alpha$ wavelength at the quasar redshift (the so-called
``Gunn-Peterson trough''), indicating a non-negligible neutral fraction in
the IGM.}
\label{fig:qsos}
\end{figure}

Several quasars beyond $z\sim6.1$ show in their spectra such a (so-called
``Gunn-Peterson'') trough, a blank spectral region at wavelengths shorter
than \lya at the quasar redshift (Figure~\ref{fig:qsos}). The detection
of Gunn-Peterson troughs indicates a rapid change in the neutral content of
the IGM at $z\sim6$, and hence a rapid change in the intensity of the
background ionizing flux. However, even a small atomic hydrogen fraction of
$\sim 10^{-3}$ would still produce nearly complete \lya absorption.

While only resonant \lya absorption is important at moderate
redshifts, the damping wing of the \lya line plays a significant role
when neutral fractions of order unity are considered at $z \gtrsim 6$.
The scattering cross-section of the \lya resonance line by neutral
hydrogen is given by
\begin{equation}
\sigma_\alpha(\nu) = {3 \lambda_\alpha^2 \Lambda_\alpha^2 \over 8\pi}
{(\nu/\nu_\alpha)^4\over
4\pi^2(\nu-\nu_\alpha)^2+(\Lambda_\alpha^2/4)(\nu/\nu_\alpha)^6}\ ,
\label{eq:sig}
\end{equation}
where $\Lambda_\alpha=(8\pi^2 e^2
f_\alpha/3m_ec\lambda_\alpha^2)=6.25\times 10^8~{\rm s^{-1}}$ is the
\lya ($2p\rightarrow 1s$) decay rate, $f_\alpha=0.4162$ is the
oscillator strength, and $\lambda_\alpha=1216$\AA\, and
$\nu_\alpha=(c/\lambda_\alpha)=2.47\times 10^{15}~{\rm Hz}$ are the
wavelength and frequency of the \lya line. The term in the numerator
is responsible for the classical Rayleigh scattering.

Although reionization is an inhomogeneous process, we consider here a
simple illustrative case of instantaneous reionization. Consider a source
at a redshift $z_s$ beyond the redshift of reionization, $\zr$, and the
corresponding scattering optical depth of a uniform, neutral IGM of
hydrogen density $n_{\rm H,0}(1+z)^3$ between the source and the
reionization redshift. The optical depth is a function of the observed
wavelength $\lambda_{\rm obs}$,
\begin{equation}
\tau(\lambda_{\rm obs})=\int_{\zr}^{z_s} dz\, {cdt\over dz}\, n_{\rm
H,0} (1+z)^3 \sigma_\alpha\left[\nu_{\rm obs}(1+z)\right]\ ,
\end{equation}
where $\nu_{\rm obs}=c/\lambda_{\rm obs}$ and
\begin{equation}
{dt\over dz} = \left[(1+z)H(z)\right]^{-1}=H_0^{-1} \times
\left[\Omega_m(1+z)^5+\Omega_\Lambda(1+z)^2+
(1-\Omega_m-\Omega_\Lambda)(1+z)^4\right]^{-1/2}\ .
\end{equation}

At wavelengths longer than \lya at the source, the optical depth
obtains a small value; these photons redshift away from the line
center along its red wing and never resonate with the line core on
their way to the observer.  Considering only the regime in which
$\vert\nu-\nu_\alpha\vert \gg \Lambda_\alpha$, we may ignore the
second term in the denominator of equation~(\ref{eq:sig}). This leads
to an analytical result for the red damping wing of the Gunn-Peterson
trough,
\begin{equation}
\tau(\lambda_{\rm obs})=\tau_s \left(\Lambda\over
4\pi^2\nu_\alpha\right) {\tilde \lambda}_{\rm obs}^{3/2}\left[
I({\tilde\lambda}_{\rm obs}^{-1}) -
I([(1+\zr)/(1+z_s)]{\tilde\lambda}_{\rm obs}^{-1})\right]\ ,
\label{eq:shift}
\end{equation}
an expression valid for ${\tilde\lambda}_{\rm obs}\geq 1$, where
$\tau_s$ is given in equation~(\ref{G-P}), and we also define
\begin{equation}
{\tilde \lambda}_{\rm obs}\equiv {\lambda_{\rm obs}\over
(1+z_s)\lambda_\alpha}
\end{equation}
and
\begin{equation}
I(x)\equiv {x^{9/2}\over 1-x}+{9\over 7}x^{7/2}+{9\over 5}x^{5/2}+ 3
x^{3/2}+9 x^{1/2}-{9\over 2} \ln\left[ {1+x^{1/2}\over 1-x^{1/2}}
\right]\ .
\end{equation}

At wavelengths shorter than 912\AA, the photons are absorbed when
they photoionize atoms of hydrogen or helium. The bound-free
absorption cross-section from the ground state of a hydrogenic ion
with nuclear charge $Z$ and an ionization threshold $h\nu_0$, is given
by 
\begin{equation}
\sigma_{bf}(\nu)= {6.30\times 10^{-18}\over Z^2}~{\rm cm^2}\times
\left({\nu_0\over
\nu}\right)^4{e^{4-(4\tan^{-1}\epsilon)/\epsilon}\over 1 -
e^{-2\pi/\epsilon}}~~~~{\rm for}~~\nu\geq\nu_0\ ,
\end{equation}
where $\epsilon\equiv \sqrt{{(\nu/\nu_0)}-1}$.  For neutral hydrogen, $Z=1$
and $\nu_{{\rm H},0}= (c/\lambda_c) =3.29\times 10^{15}$ Hz ($h\nu_{\rm
H,0}=13.60$ eV); for singly-ionized helium, $Z=2$ and $\nu_{\rm He~II, 0}=
1.31\times 10^{16}~{\rm Hz}$ ($h\nu_{\rm He~II, 0}=54.42$ eV). The
cross-section for neutral helium is more complicated; when averaged over
its narrow resonances it can be fitted to an accuracy of a few percent up
to $h\nu=50$ keV by the fitting function 
\begin{equation} 
\sigma_{bf,{\rm
He~I}}(\nu)=9.492\times 10^{-16}~{\rm cm^2}\, \times
\left[(x-1)^2+4.158\right]y^{-1.953}\left(1+ 0.825
y^{1/4}\right)^{-3.188}\ , 
\end{equation} 
where $x\equiv[(\nu/3.286\times 10^{15}~{\rm Hz}) -0.4434]$, $y\equiv
x^2+4.563$, and the threshold for ionization is $\nu_{\rm He~I,
0}=5.938\times 10^{15}~{\rm Hz}$ ($h\nu_{\rm He~I,0}=24.59$ eV).

For rough estimates, the average photoionization cross-section for a
mixture of hydrogen and helium with cosmic abundances can be
approximated in the range of $54<h\nu \lesssim 10^3$ eV as
$\sigma_{bf}\approx \sigma_0 (\nu/\nu_{\rm H,0})^{-3}$, where
$\sigma_0\approx 6\times 10^{-17}~{\rm cm^2}$.
The redshift factor in the cross-section then cancels exactly the
redshift evolution of the gas density and the resulting optical depth
depends only on the elapsed cosmic time, $t(\zr)-t(z_s)$. At high
redshifts this yields
\begin{eqnarray}
\tau_{bf}(\lambda_{\rm obs})&=&\int_{\zr}^{z_s} dz {cdt\over dz}
n_{0} (1+z)^3 \sigma_{\rm bf}\left[\nu_{\rm obs}(1+z)\right]
\nonumber \\ &\approx&
1.5\times 10^2 \left({\Omega_bh\over 0.03}\right) \left({\Omega_m\over
  0.3}\right)^{-1/2}\left({\lambda\over 100{\mbox{\AA}}}\right)^{3}\times
 \left[{1\over (1+\zr)^{3/2}}
-{1\over (1+z_s)^{3/2}}\right]\ . \label{eq:bf}
\end{eqnarray}
The bound-free optical depth only becomes of order unity in the
extreme ultraviolet (UV) to soft X-rays, around $h\nu \sim 0.1$ keV, a
regime which is unfortunately difficult to observe due to Galactic
absorption.

\subsection{21-cm absorption or emission}
\label{sec:21-cmAtomic}

\subsubsection{The spin temperature of the 21-cm transition of hydrogen}

The ground state of hydrogen exhibits hyperfine splitting owing to the
possibility of two relative alignments of the spins of the proton and the
electron. The state with parallel spins (the triplet state) has a slightly
higher energy than the state with anti-parallel spins (the singlet
state). The 21-cm line associated with the spin-flip transition from the
triplet to the singlet state is often used to detect neutral hydrogen in
the local universe. At high redshift, the occurrence of a neutral
pre-reionization IGM offers the prospect of detecting the first sources of
radiation and probing the reionization era by mapping the 21-cm emission
from neutral regions. While its energy density is estimated to be only a
$1\%$ correction to that of the CMB, the redshifted 21-cm emission should
display angular structure as well as frequency structure due to
inhomogeneities in the gas density field, hydrogen ionized fraction, and
spin temperature. Indeed, a full mapping of the distribution of H~I as a
function of redshift is possible in principle.

The basic physics of the hydrogen spin transition is determined as
follows. The ground-state hyperfine levels of hydrogen tend to thermalize
with the CMB background, making the IGM unobservable. If other processes
shift the hyperfine level populations away from thermal equilibrium, then
the gas becomes observable against the CMB in emission or in
absorption. The relative occupancy of the spin levels is usually described
in terms of the hydrogen spin temperature $T_S$, defined by \beq
\frac{n_1}{n_0}=3\, \exp\left\{-\frac{T_*}{T_S}\right\}\ , \eeq where $n_0$
and $n_1$ refer respectively to the singlet and triplet hyperfine levels in
the atomic ground state ($n=1$), and $T_*=0.068$ K is defined by $k_B
T_*=E_{21}$, where the energy of the 21 cm transition is $E_{21}=5.9 \times
10^{-6}$ eV, corresponding to a frequency of 1420 MHz. In the presence of
the CMB alone, the spin states reach thermal equilibrium with $T_S=T_{\rm
CMB}=2.725 (1+z)$ K on a time-scale of $T_*/(T_{\rm CMB} A_{10}) \simeq 3
\times 10^5 (1+z)^{-1}$ yr, where $A_{10}=2.87 \times 10^{-15}$ s$^{-1}$ is
the spontaneous decay rate of the hyperfine transition. This time-scale is
much shorter than the age of the universe at all redshifts after
cosmological recombination.

The IGM is observable when the kinetic temperature $T_k$ of the gas differs
from $T_{\rm CMB}$ and an effective mechanism couples $T_S$ to
$T_k$. Collisional de-excitation of the triplet level dominates at very
high redshift, when the gas density (and thus the collision rate) is still
high, but once a significant galaxy population forms in the universe, the
spin temperature is affected also by an indirect mechanism that acts
through the scattering of Lyman-$\alpha$ photons. Continuum UV photons
produced by early radiation sources redshift by the Hubble expansion into
the local Lyman-$\alpha$ line at a lower redshift. These photons mix the
spin states via the Wouthuysen-Field process whereby an atom initially in
the $n=1$ state absorbs a Lyman-$\alpha$ photon, and the spontaneous decay
which returns it from $n=2$ to $n=1$ can result in a final spin state which
is different from the initial one. Since the neutral IGM is highly opaque
to resonant scattering, and the Lyman-$\alpha$ photons receive Doppler
kicks in each scattering, the shape of the radiation spectrum near
Lyman-$\alpha$ is determined by $T_k$ (Field 1959), and the resulting spin
temperature (assuming $T_S \gg T_*$) is then a weighted average of $T_k$
and $T_{\rm CMB}$: \beq T_S=\frac{T_{\rm CMB} T_k (1+x_{\rm tot}) }{T_k +
T_{\rm CMB} x_{\rm tot}}\ , \eeq where $x_{\rm tot} = x_{\alpha} + x_c$ is
the sum of the radiative and collisional threshold parameters. These
parameters are
\begin{equation}
x_{\alpha} = {{P_{10} T_\star}\over {A_{10} T_{\rm CMB}}}\ ,
\end{equation}
and
\begin{equation}
x_c = {{4 \kappa_{1-0}(T_k)\, n_H T_\star}\over {3 A_{10} T_{\rm CMB}}}\ ,\
\end{equation} where $P_{10}$ is the indirect de-excitation rate of the
triplet $n=1$ state via the Wouthuysen-Field process, related to the total
scattering rate $P_{\alpha}$ of Lyman-$\alpha$ photons by $P_{10}=4
P_{\alpha}/27$. Also, the atomic coefficient $\kappa_{1-0}(T_k)$ is
tabulated as a function of $T_k$. The coupling of the spin temperature to
the gas temperature becomes substantial when $x_{\rm tot} \gtrsim 1$; in
particular, $x_{\alpha} = 1$ defines the thermalization rate of
$P_{\alpha}$: \beq P_{\rm th} \equiv \frac{27 A_{10} T_{\rm CMB}}{4 T_*}
\simeq 7.6 \times 10^{-12}\, \left(\frac{1+z}{10}\right)\ {\rm s}^{-1}\
. \eeq

A patch of neutral hydrogen at the mean density and with a uniform
$T_S$ produces (after correcting for stimulated emission) an optical
depth at a present-day (observed) wavelength of $21 (1+z)$ cm,
\beq \tau(z) = 9.0
\times 10^{-3} \left(\frac{T_{\rm CMB}} {T_S} \right) \left (
\frac{\Omega_b h} {0.03} \right) \left(\frac{\Omm}{0.3}\right)^ {-1/2}
\left(\frac{1+z}{10}\right)^{1/2}\ , \eeq assuming a high redshift
$z\gg1$. The observed spectral intensity $I_{\nu}$ relative to the CMB at a
frequency $\nu$ is measured by radio astronomers as an effective
brightness temperature $T_b$ of blackbody emission at this frequency,
defined using the Rayleigh-Jeans limit of the Planck radiation
formula: $I_{\nu} \equiv 2 k_B T_b \nu^2 / c^2 $.

The brightness temperature through the IGM is $T_b=T_{\rm CMB}
e^{-\tau}+T_S (1-e^{-\tau})$, so the observed differential antenna
temperature of this region relative to the CMB is \beqa T_b&=&(1+z)^{-1}
(T_S-T_{\rm CMB}) (1-e^{-\tau}) \nonumber \\ &\simeq& 28\, {\rm mK}\,
\left( \frac{\Omega_b h} {0.033} \right) \left(\frac{\Omm}{0.27}\right)^
{-1/2} \left( \frac{1+z} {10} \right)^{1/2} \left( \frac{T_S-T_{\rm CMB}}
{T_S} \right)\ , \eeqa where $\tau \ll 1$ is assumed and $T_b$ has been
redshifted to redshift zero. Note that the combination that appears in
$T_b$ is
\begin{equation}
{T_S - T_{\rm CMB} \over T_S} = {x_{\rm tot}\over 1+ x_{\rm tot}}
\left(1 - {T_{\rm CMB}\over T_k} \right)\ .
\end{equation}
In overdense regions, the observed $T_b$ is proportional to the
overdensity, and in partially ionized regions $T_b$ is proportional to the
neutral fraction. Also, if $T_S \gg T_{\rm CMB}$ then the IGM is observed
in emission at a level that is independent of $T_S$. On the other hand, if
$T_S \ll T_{\rm CMB}$ then the IGM is observed in absorption at a level
that is enhanced by a factor of $T_{\rm CMB} / T_S$. As a result, a number
of cosmic events are expected to leave observable signatures in the
redshifted 21-cm line, as discussed below in further detail.

\begin{figure}
\epsfxsize=8cm \epsfbox{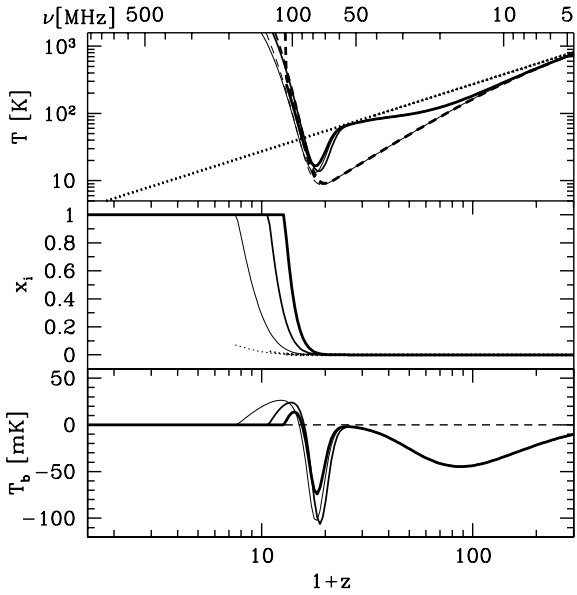}
\caption{{\em Top panel: }Evolution with redshift $z$ of the CMB
temperature $T_{\rm CMB}$ (dotted curve),the gas kinetic temperature $T_k$
(dashed curve), and the spin temperature $T_S$ (solid curve), taken from
Pritchard \& Loeb (2008).  {\em Middle panel: }Evolution of the gas
fraction in ionized regions $x_i$ (solid curve) and the ionized fraction
outside these regions (due to diffuse X-rays) $x_e$ (dotted curve). {\em
Bottom panel: } Evolution of mean 21 cm brightness temperature $T_b$.  The
horizontal axis at the top provides the observed photon frequency at the
different redshifts shown at the bottom.  Each panel shows curves for three
models in which reionization is completed at different redshifts, namely
$z=6.47$ (thin curves), $z=9.76$ (medium curves), and $z=11.76$ (thick
curves).  }
\label{fig:tempZlong}
\end{figure}

Figure \ref{fig:tempZlong} illustrates the mean IGM evolution for three
examples in which reionization is completed at different redshifts, namely
$z=6.47$ (thin curves), $z=9.76$ (medium curves), and $z=11.76$ (thick
curves).  The top panel shows the global evolution of the CMB temperature
$T_{\rm CMB}$ (dotted curve), the gas kinetic temperature $T_k$ (dashed
curve), and the spin temperature $T_S$ (solid curve).  The middle panel
shows the evolution of the ionized gas fraction and the bottom panel
presents the mean 21 cm brightness temperature, $T_b$.

\subsubsection{A handy tool for studying cosmic reionization}

The prospect of studying reionization by mapping the distribution of atomic
hydrogen across the universe using its prominent 21-cm spectral line has
motivated several teams to design and construct arrays of low-frequency
radio telescopes; the Low Frequency Array (http://www.lofar.org/), the
Mileura Wide-Field Array ({\it
http://www.haystack.mit.edu/ast/arrays/mwa/site/index.html}), the Primeval
Structure Telescope ({\it http://arxiv.org/abs/astro-ph/0502029}), and
ultimately the Square Kilometer Array ({\it http://www.skatelescope.org})
will search over the next decade for 21-cm emission or absorption from
$z\sim 6.5$--15, redshifted and observed today at relatively low
frequencies which correspond to wavelengths of 1.5 to 4 meters.

The idea is to use the resonance associated with the hyperfine splitting in
the ground state of hydrogen. While the CMB spectrum peaks at a wavelength
of 2 mm, it provides a still-measurable intensity at meter wavelengths that
can be used as the bright background source against which we can see the
expected 1\% absorption by neutral hydrogen along the line of sight. The
hydrogen gas produces 21-cm absorption if its spin temperature is colder
than the CMB and excess emission if it is hotter. Since the CMB covers the
entire sky, a complete three-dimensional map of neutral hydrogen can in
principle be made from the sky position of each absorbing gas cloud
together with its redshift $z$.  Different observed wavelengths slice the
Universe at different redshifts, and ionized regions are expected to appear
as cavities in the hydrogen distribution, similar to holes in swiss cheese.
Because the smallest angular size resolvable by a telescope is proportional
to the observed wavelength, radio astronomy at wavelengths as large as a
meter has remained relatively undeveloped. Producing resolved images even
of large sources such as cosmological ionized bubbles requires telescopes
which have a kilometer scale. It is much more cost-effective to use a large
array of thousands of simple antennas distributed over several kilometers,
and to use computers to cross-correlate the measurements of the individual
antennas and combine them effectively into a single large telescope. The
new experiments are being placed mostly in remote sites, because the cosmic
wavelength region overlaps with more mundane terrestrial
telecommunications.

\begin{figure}
\epsfxsize=16cm \epsfbox{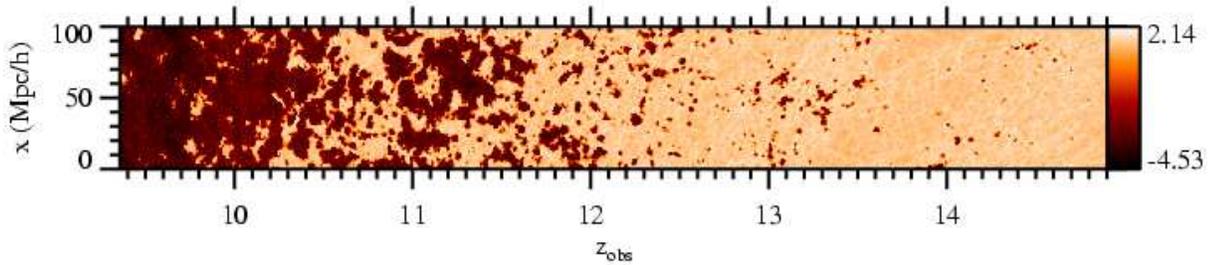}
\caption{Close-up of cosmic evolution during the epoch of reionization, as
revealed in a predicted 21-cm map of the IGM based on a numerical
simulation (from Mellema et al. 2006). This map is constructed from slices
of the simulated cubic box of side 150 Mpc (in comoving units), taken at
various times during reionization, which for the parameters of this
particular simulation spans a period of 250 million years from redshift 15
down to 9.3. The vertical axis shows position $\chi$ in units of Mpc/h
(where $h=0.7$). This two-dimensional slice of the sky (one linear
direction on the sky versus the line-of-sight or redshift direction) shows
$\log_{10}(T_b)$, where $T_b$ (in mK) is the 21-cm brightness temperature
relative to the CMB. Since neutral regions correspond to strong emission
(i.e., a high $T_b$), this slice illustrates the global progress of
reionization and the substantial large-scale spatial fluctuations in
reionization history. Observationally it corresponds to a narrow strip half
a degree in length on the sky observed with radio telescopes over a
wavelength range of 2.2 to 3.4 m (with each wavelength corresponding to
21-cm emission at a specific redshift slice).}
\label{fig:Mellema}
\end{figure}

\begin{figure}
\epsfxsize=10cm \epsfbox{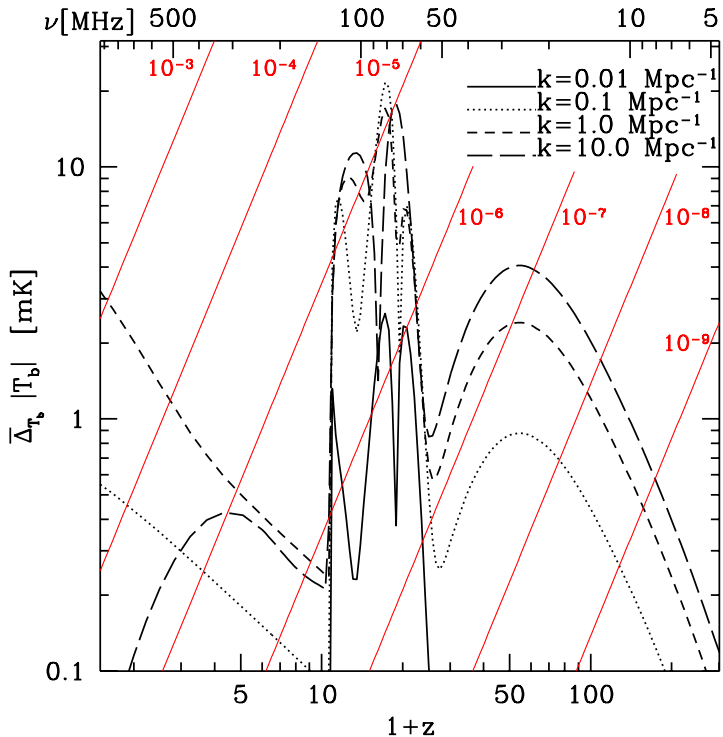}
\caption{Predicted redshift evolution of the angle-averaged amplitude of
the 21-cm power spectrum ($|\bar{\Delta}_{T_b}|=[k^3P_{\rm
21-cm}(k)/2\pi^2]^{1/2}$) at comoving wavenumbers $k=0.01$ (solid curve),
0.1 (dotted curve), 1.0 (short dashed curve), 10.0 (long dashed curve), and
100.0${\rm Mpc}^{-1}$ (dot-dashed curve). In the model shown, reionization
is completed at $z=9.76$.  The horizontal axis at the top shows the
observed photon frequency at the different redshifts.  The diagonal
straight (red) lines show various factors of suppression for the
synchrotron Galactic foreground, necessary to reveal the 21-cm signal (from
Pritchard \& Loeb 2008).}
\label{fig:Analytic}
\end{figure}

In approaching redshifted 21-cm observations, although the first inkling
might be to consider the mean emission signal in the bottom panel of
Figure~\ref{fig:tempZlong}, the signal is orders of magnitude fainter than
foreground synchrotron emission from relativistic electrons in the magnetic
field of our own Milky Way as well as other galaxies (see
Figure~\ref{fig:Analytic}). Thus cosmologists have focused on the expected
characteristic variations in $T_b$, both with position on the sky and
especially with frequency, which signifies redshift for the cosmic
signal. The synchrotron foreground is expected to have a smooth frequency
spectrum, and so it is possible to isolate the cosmological signal by
taking the difference in the sky brightness fluctuations at slightly
different frequencies (as long as the frequency separation corresponds to
the characteristic size of ionized bubbles). The 21-cm brightness
temperature depends on the density of neutral hydrogen. As explained in the
previous subsection, large-scale patterns in the reionization are driven by
spatial variations in the abundance of galaxies; the 21-cm fluctuations
reach $\sim$5 mK (root mean square) in brightness temperature
(Figure~\ref{fig:Mellema}) on a scale of 10 Mpc (comoving). While detailed
maps will be difficult to extract due to the foreground emission, a
statistical detection of these fluctuations (through the power spectrum) is
expected to be well within the capabilities of the first-generation
experiments now being built. Current work suggests that the key information
on the topology and timing of reionization can be extracted statistically.

\begin{figure}
\epsfxsize=12cm \epsfbox{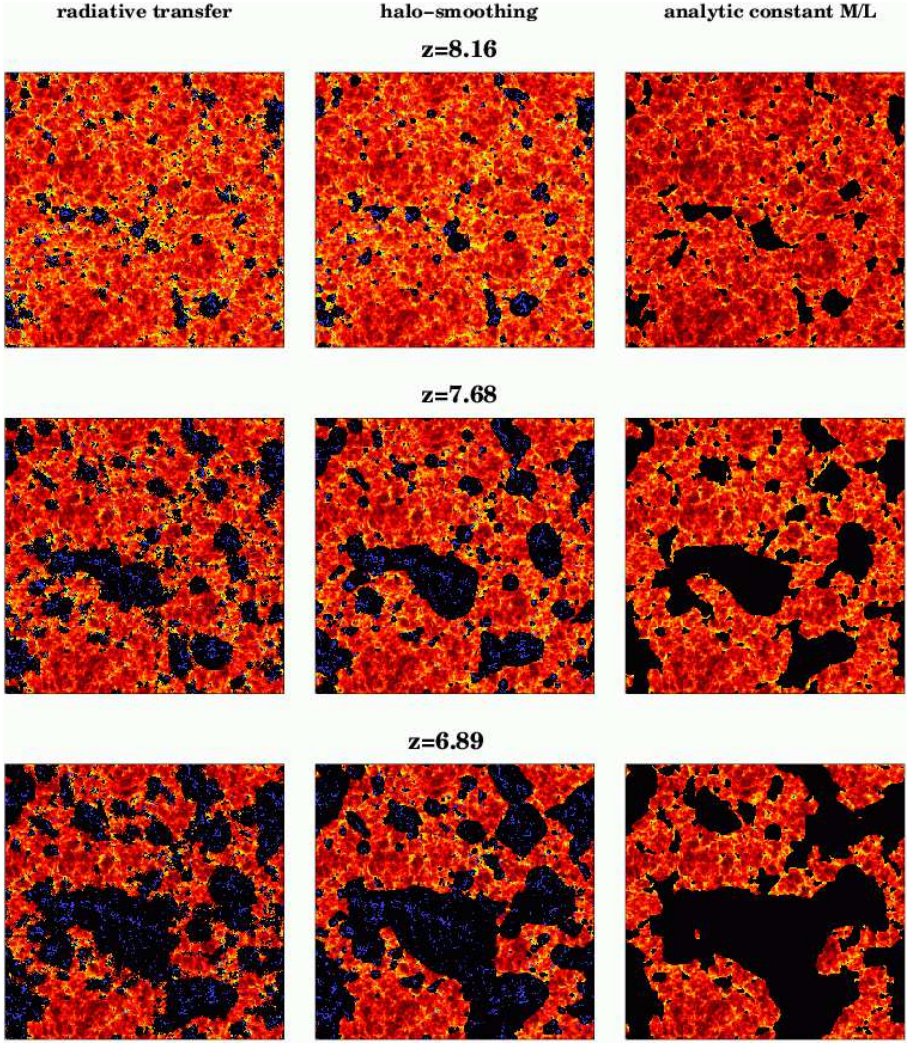}
\caption{Maps of the 21-cm brightness temperature comparing results of a
numerical simulation and of two simpler numerical schemes, at three
different redshifts (from Zahn et al. 2006). Each map is 65.6 Mpc/$h$ on a
side, with a depth (0.25 Mpc/$h$) that is comparable to the frequency
resolution of planned experiments. The ionized fractions are $x_{\rm
i}=0.13$, 0.35 and 0.55 for $z=8.16$, 7.26 and 6.89 (top to bottom),
respectively. All three maps show a very similar large-scale ionization
topology. \emph{Left column:} Numerical simulation, showing the ionized
bubbles (black) produced by the ionizing sources (blue dots) that form in
the simulation. \emph{Middle column:} Numerical scheme that applies an
analytical model to the final distribution of ionizing sources that form in
the simulation. \emph{Right column:} Numerical scheme that applies the
analytical model to the linear density fluctuations that are the initial
conditions of the simulation.}
\label{fig:Zahn}
\end{figure}

While numerical simulations of reionization are now reaching the
cosmological box sizes needed to predict the large-scale topology of the
ionized bubbles, they do this at the price of limited small-scale
resolution (see Figure \ref{fig:Zahn}). These simulations cannot yet follow
in any detail the formation of individual stars within galaxies, or the
feedback that stars produce on the surrounding gas, such as photo-heating
or the hydrodynamic and chemical impact of supernovae, which blow hot
bubbles of gas enriched with the chemical products of stellar
nucleosynthesis. Thus, the simulations cannot directly predict whether the
stars that form during reionization are similar to the stars in the Milky
Way and nearby galaxies or to the primordial $100 M_{\odot}$ stars. They
also cannot determine whether feedback prevents low-mass dark matter halos
from forming stars. Thus, models are needed that make it possible to vary
all these astrophysical parameters of the ionizing sources and to study the
effect on the 21-cm observations.

The theoretical expectations presented here for reionization and for the
21-cm signal are based on rather large extrapolations from observed
galaxies to deduce the properties of much smaller galaxies that formed at
an earlier cosmic epoch. Considerable surprises are thus possible, such as
an early population of quasars or even unstable exotic particles that
emitted ionizing radiation as they decayed. In any case, the forthcoming
observational data in 21-cm cosmology should make the next few years a very
exciting time.

At high redshifts prior to reionization, spatial perturbations in the
thermodynamic gas properties are linear and can be predicted precisely (see
section~\ref{sec:lin}). Thus, if the gas is probed with the 21-cm technique
then it becomes a promising tool of fundamental, precision cosmology, able
to probe the primordial power spectrum of density fluctuations imprinted in
the very early universe, perhaps in an era of cosmic inflation. The 21-cm
fluctuations can be measured down to the smallest scales where the baryon
pressure suppresses gas fluctuations, while the CMB anisotropies are damped
on small scales (through the so-called Silk damping). This difference in
damping scales can be seen by comparing the baryon-density and
photon-temperature power spectra in Figure~\ref{fig:photons}.  Since the
21-cm technique is also three-dimensional (while the CMB yields a single
sky map), there is a much large potential number of independent modes
probed by the 21-cm signal: $N_{\rm 21-cm}\sim 3 \times 10^{16}$ compared
to $N_{\rm cmb} \sim 2\times 10^7$. This larger number should provide a
measure of non-Gaussian deviations to a level of $\sim N_{\rm 21
cm}^{-1/2}$, constituting a test of the inflationary origin of the
primordial inhomogeneities which are expected to possess non-Gaussian
deviations $\gtrsim 10^{-6}$. 

The 21cm fluctuations are expected to simply trace the primordial
power-spectrum of matter density perturbations (which is shaped by the
initial conditions from inflation and the dark matter) either before the
first population of galaxies had formed (at redshifts $z>25$) or after
reionization ($z<6$) -- when only dense pockets of self-shielded hydrogen
(such as damped Ly$\alpha$ systems) survive.  During the epoch of
reionization, the fluctuations are mainly shaped by the topology of ionized
regions, and thus depend on uncertain astrophysical details involving star
formation.  However, even during this epoch, the imprint of peculiar
velocities (which are induced gravitationally by density fluctuations), can
in principle be used to separate the implications for fundamental physics
from the astrophysics.

\begin{figure}
\epsfxsize=10cm \epsfbox{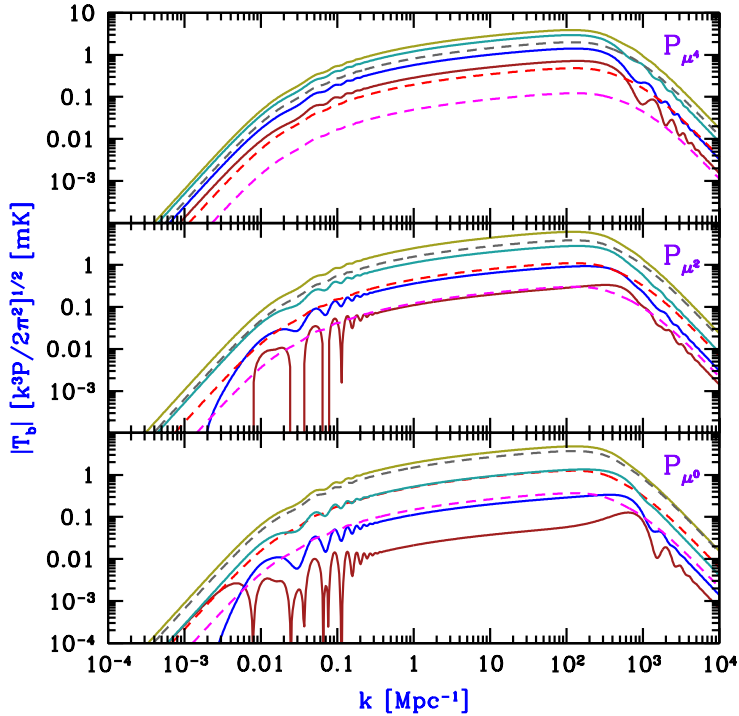}
\caption{Power spectra of 21-cm brightness fluctuations versus comoving
wavenumber (from Barkana \& Loeb 2005c). Show are the three power spectra
that are separately observable, $P_{\mu^4}$ (upper panel), $P_{\mu^2}$
(middle panel), and $P_{\mu^0}$ (lower panel). Each case shows redshifts
200, 150, 100, 50 (solid curves, from bottom to top), 35, 25, and 20
(dashed curves, from top to bottom).}
\label{21-cmP}
\end{figure}

Peculiar velocities imprint a particular form of anisotropy in
the 21-cm fluctuations that is caused by gas motions along the line of
sight. This anisotropy, expected in any measurement of density that is
based on a spectral resonance or on redshift measurements, results from
velocity compression. Consider a photon traveling along the line of sight
that resonates with absorbing atoms at a particular point. In a uniform,
expanding universe, the absorption optical depth encountered by this photon
probes only a narrow strip of atoms, since the expansion of the universe
makes all other atoms move with a relative velocity that takes them outside
the narrow frequency width of the resonance line. If there is a density
peak, however, near the resonating position, the increased gravity will
reduce the expansion velocities around this point and bring more gas into
the resonating velocity width. This effect is sensitive only to the
line-of-sight component of the velocity gradient of the gas, and thus
causes an observed anisotropy in the power spectrum even when all physical
causes of the fluctuations are statistically isotropic. This anisotropy is
particularly important in the case of 21-cm fluctuations. When all
fluctuations are linear, the 21-cm power spectrum takes the form \beq
P_{\rm 21-cm}({\bf k}) = \mu^4 P_{\rho}(k) + 2 \mu^2 P_{\rho - {\rm iso}}
(k) + P_{\rm iso}\ , \eeq where $\mu = \cos \theta$ in terms of the angle
$\theta$ between the wave-vector ${\bf k}$ of a given Fourier mode and the
line of sight, $P_{\rm iso}$ is the isotropic power spectrum that would
result from all sources of 21-cm fluctuations without velocity compression,
$P_{\rho}(k)$ is the 21-cm power spectrum from gas density fluctuations
alone, and $P_{\rho - {\rm iso}} (k)$ is the Fourier transform of the
cross-correlation between the density and all sources of 21-cm
fluctuations. The three power spectra can also be denoted $P_{\mu^4}(k)$,
$P_{\mu^2}(k)$, and $P_{\mu^0}(k)$, according to the power of $\mu$ that
multiplies each term.  The prediction for these power spectra at high
redshift ($z > 20$), neglecting the effects of any stellar radiation, are
shown in Figure~\ref{21-cmP}.  At these redshifts, the 21-cm fluctuations
probe the infall of the baryons into the dark matter potential wells. The
power spectrum shows remnants of the photon-baryon acoustic oscillations on
large scales, and of the baryon pressure suppression on small scales.

Once stellar radiation becomes significant, many processes can contribute
to the 21-cm fluctuations. The contributions include fluctuations in gas
density, temperature, ionized fraction, and \Lya flux. These processes can
be divided into two broad categories: The first, related to {\it
``physics''}, consists of probes of fundamental, precision cosmology, and
the second, related to {\it ``astrophysics''}, consists of probes of
stars. Both categories are interesting -- the first for precision measures
of cosmological parameters and studies of processes in the early universe,
and the second for studies of the properties of the first
galaxies. However, the astrophysics depends on complex non-linear processes
(collapse of dark matter halos, star formation, supernova feedback), and
must be cleanly separated from the physics contribution, in order to allow
precision measurements of the latter.  As long as all the fluctuations are
linear, the anisotropy noted above allows precisely this separation of the
{\it fundamental physics} from the {\it astrophysics} of the 21-cm
fluctuations. In particular, the $P_{\mu^4}(k)$ is independent of the
effects of stellar radiation, and is a clean probe of the gas density
fluctuations. Once non-linear terms become important, there arises a
significant mixing of the different terms; in particular, this occurs on
the scale of the ionizing bubbles during reionization.

The 21-cm fluctuations are affected by fluctuations in the Lyman-$\alpha$
flux from stars, a result that yields an indirect method to detect and
study the early population of galaxies at $z \sim 20$. The fluctuations are
caused by biased inhomogeneities in the density of galaxies, along with
Poisson fluctuations in the number of galaxies. Observing the power-spectra
of these two sources would probe the number density of the earliest
galaxies and the typical mass of their host dark matter halos. Furthermore,
the enhanced amplitude of the 21-cm fluctuations from the era of \Lya
coupling improves considerably the practical prospects for their
detection. Precise predictions account for the detailed properties of all
possible cascades of a hydrogen atom after it absorbs a photon. Around the
same time, X-rays may also start to heat the cosmic gas, producing strong
21-cm fluctuations due to fluctuations in the X-ray flux.

\section{Conclusions}

The initial conditions of our Universe can be summarized on a single sheet
of paper. Yet the Universe is full of complex structures today, such as
stars, galaxies and groups of galaxies.  This review discussed the standard
theoretical model for how complexity emerged from the simple initial state
of the Universe through the action of gravity. In order to test and inform
the related theoretical calculations, large-aperture telescopes and arrays
of radio antennae are currently being designed and constructed.

The actual transition from simplicity to complexity has not been observed
as of yet.  The simple initial conditions were already traced in maps of
the microwave background radiation, but the challenge of detecting the
first generation of galaxies defines one of the exciting frontiers in the
future of cosmology.  Once at hand, the missing images of the infant
Universe might potentially surprise us and revise our current ideas.

\subsection*{Acknowledgements}

I thank my collaborators on the topics covered by this review: Dan Babich,
Rennan Barkana, Volker Bromm, Steve Furlanetto, Zoltan Haiman, Jonathan
Pritchard, Stuart Wyithe, and Matias Zaldarriaga.

\noindent

\subsection*{Bibliography}

\Ref {Abel T.L., Bryan G.L., Norman M.L. (2002). The Formation of the First Star in the Universe. {\it Science} \textbf{295} 93. [Results from 
simulations of the formation of the first stars]. }

\Ref {Allison A.C. and Dalgarno A. (1969). Spin Change in Collisions of Hydrogen Atoms. {\it Astrophys. J.} \textbf{158} 423. [An early paper
about the effect of collisions on the spin temperature of hydrogen].}

\Ref {Arons J. and Wingert D.W. (1972). Theoretical Models of Photoionized Intergalactic Hydrogen. {\it Astrophys. J.} \textbf{177} 1. [An early discussion 
on the re-ionization of the intergalactic hydrogen, more
than a quarter of a century before th topic gained popularity]. }

\Ref {Babich D. and Loeb A. (2006). Imprint of Inhomogeneous Reionization on the Power Spectrum of Galaxy Surveys at High Redshifts. {\it Astrophys. J.} \textbf{640} 1. [A discussion on the imprint of cosmic reionization 
on the distribution of low-mass galaxies].}

\Ref {Barkana R. and Loeb A. (2001). In the beginning: the first sources of light and the reionization of the universe. {\it Phys. Rep.} \textbf{349} 125.
[An overview on the physics of the first galaxies and reionization].}

\Ref  { Barkana R. and Loeb A. (2004a). Gamma-Ray Bursts versus Quasars: Ly$\alpha$ Signatures of Reionization versus Cosmological Infall. {\it Astrophys. J.} \textbf{601} 64. [A comparison between quasars and gamma-ray bursts 
as probes of the intergalactic medium at high redshifts].}

\Ref { Barkana R. and Loeb A. (2004b).  Unusually Large Fluctuations in the
Statistics of Galaxy Formation at High Redshift. {\it Astrophys. J.}
\textbf{609} 474. [A discussion on the limitations of computer simulations
of reionization owing to their finite size of the simulated region]. }

\Ref { Barkana R. and Loeb A. (2005a). A Method for Separating the Physics
from the Astrophysics of High-Redshift 21 Centimeter Fluctuations. {\it
Astrophys. J. Lett.} \textbf{624} L65. [The imprint of peculiar velocities
on the 21cm brightness fluctuations can be used to extract cosmological
information during the epoch of reionization]. }

\Ref  { Barkana R. and Loeb A. (2005b). Detecting the Earliest Galaxies through Two New Sources of 21 Centimeter Fluctuations. {\it Astrophys. J.}
\textbf{626} 1. [The imprint of the first galaxies on the 21cm
fluctuations. }

\Ref  { Barkana R. and Loeb A. (2005c). Probing the epoch of early baryonic infall through 21-cm fluctuations. {\it Mon. Not. Roy. Astron. Soc.
Lett.}  \textbf{363} L36. [A discussion on the signature of acoustic 
oscillations on the 21cm brightness fluctuations]. }

\Ref { Barkana R. and Loeb A. (2007). The physics and early history of the
intergalactic medium. {\it Rep.  Prog. Phys.} \textbf{70} 627.
[A review on the history of the intergalactic medium].}

\Ref { Bennett C.L. et al. (1996). Four-Year COBE DMR Cosmic Microwave
Background Observations: Maps and Basic Results. {\it Astrophys. J. Lett.}
\textbf{464} L1. [A description of the first robust detection of microwave
background fluctuations by the COBE satellite (for which the Nobel prize
was awarded in 2006].}

\Ref  { Bharadwaj S. and Ali S.S. (2004). The cosmic microwave background radiation fluctuations from HI perturbations prior to reionization. {\it Mon. Not. Roy. Astron. Soc.} \textbf{352} 142. [A discussion on the imprint of peculiar 
velocities on 21cm fluctuations].}

\Ref  { Bowman J.D., Morales M.F. and Hewitt J.N. (2006). The Sensitivity of First-Generation Epoch of Reionization Observatories and Their Potential for Differentiating Theoretical Power Spectra. {\it Astrophys. J.} \textbf{638} 20.
[An early discussion on the feasibility of modern measurements of the 21cm fluctuations from the epoch of reionization].}

\Ref  { Bromm V., Coppi P.S., Larson R.B. (2002). The Formation of the First Stars. I. The Primordial Star-forming Cloud. {\it Astrophys. J.} \textbf{564} 23.
[Results from numerical simulations of the formation of the first stars].}

\Ref  { Bromm V. and Larson R.B. (2004). The First Stars. {\it Ann. Rev. Astron. \& Astrophys.} \textbf{42} 79. [An overview on the formation of the 
the first stars].}

\Ref { Bromm V. and Loeb A. (2003).  Formation of the First Supermassive
Black Holes. {\it Astrophys. J.}  \textbf{596} 34. [An early model for the
production of massive seeds for quasar black holes at early cosmic times].}

\Ref { Bromm V. and Loeb A. (2004). Accretion onto a primordial
protostar. {\it New Astronomy} \textbf{9} 353. [Results from high-resultion
simulations of the first stars, including an estimate of their final
mass].}

\Ref { Bromm V., Kudritzki R.P. and Loeb A. (2001). Generic Spectrum and
Ionization Efficiency of a Heavy Initial Mass Function for the First Stars.
{\it Astrophys. J.} \textbf{552} 464. [A pioneering derivation of the
spectrum of the first stars, and the number of ionizing photons they
produce per stellar mass].}

\Ref { Couchman, H.M.P. and Rees M.J. (1986). Pregalactic evolution in
cosmologies with cold dark matter. {\it Mon. Not. Roy. Astron. Soc.}
\textbf{221} 53. [A pioneering paper on the formation of the first
galaxies in a CDM cosmology].}

\Ref { Chen X. and Miralda-Escud\'e J. (2004). The Spin-Kinetic Temperature
Coupling and the Heating Rate due to Ly$\alpha$ Scattering before
Reionization: Predictions for 21 Centimeter Emission and Absorption. {\it
Astrophys. J.} \textbf{602} 1. [A detailed calculation of the effect of
Lyman-$\alpha$ photons on the spin temperature of intergalactic hydrogen].}

\Ref { Ciardi B., Ferrara A. and White S.D.M. (2003). Early reionization by
the first galaxies. {\it Mon. Not. Roy. Astron. Soc.} \textbf{344} L7.
[Results from simulations of reionization].}

\Ref { Ciardi B. and Loeb A. (2000).  Expected Number and Flux Distribution
of Gamma-Ray Burst Afterglows with High Redshifts.  {\it Astrophys. J.}
\textbf{540} 687. [An early calculation of the rate of gamma-ray bursts
eith high-redshitfs].}
 
\Ref { Cole S. et al. (2005). The 2dF Galaxy Redshift Survey:
power-spectrum analysis of the final data set and cosmological
implications.  {\it Mon. Not. R. Astron. Soc.} \textbf{362} 505. [Recent
data on the distribution of galaxies on large spatial scales].}

\Ref { Dijkstra M., Haiman Z., Rees M.J. and Weinberg
D.H. (2004). Photoionization Feedback in Low-Mass Galaxies at High
Redshift. {\it Astrophys. J.} \textbf{601} 666. [A recent discussion on the
suppression of low-mass galaxies after reionization].}

\Ref { Di Matteo T., Perna R., Abel T. and Rees M.J. (2002). Radio
Foregrounds for the 21 Centimeter Tomography of the Neutral Intergalactic
Medium at High Redshifts. {\it Astrophys. J.} \textbf{564} 576. [A
discussion on the contaminating noise for future 21cm observations].}

\Ref {Di Matteo T., Springel V., \& Hernquist, L. (2005). Energy input from
quasars regulates the growth and activity of black holes and their host
galaxies. {\it Nature}, \textbf{433}, 604. [Simulations of quasar feedback
on galaxy formation and evolution].}

\Ref { Eisenstein D.J. et al. (2005). Detection of the Baryon Acoustic Peak
in the Large-Scale Correlation Function of SDSS Luminous Red Galaxies. {\it
Astrophys. J.} \textbf{633} 560. [The first detection of baryonic
oscillations in the distribution of galaxies].}

\Ref  { Efstathiou G. (1992).  Suppressing the formation of dwarf galaxies via photoionization. {\it Mon. Not. Roy Astron. Soc.}  \textbf{256} 43.
[An early discussion on the suppression of dwarf galaxies
by ionizing radiation].}

\Ref { Ellis R. (2008). Observations of the High Redshift Universe.  {\it
SAAS-Fee Advanced Course 36}, Springer Verlag, Berlin
2008. http://arxiv.org/abs/astro-ph/0701024.  [A recent overview of the
status of observations of high-redshift galaxies].}

\Ref { Fan X. et al. (2002). Evolution of the Ionizing Background and the
Epoch of Reionization from the Spectra of $ z \sim 6$ Quasars. {\it
Astron. J.} \textbf{123} 1247. [Observational constraints on intergalactic
hydrogen from the spectra of quasars which formed a billion years after the
big bang].}

\Ref { Fan X. et al. (2003). A Survey of $z>5.7$ Quasars in the Sloan
Digital Sky Survey. II. Discovery of Three Additional Quasars at $
z>6$. {\it Astron. J.} \textbf{125} 1649. [Observations of high redshift
quasars].}

\Ref { Fan X. et al. (2005).  Constraining the Evolution of the Ionizing
Background and the Epoch of Reionization with $ z \sim 6$ Quasars II: A
Sample of 19 Quasars. {\it Astron. J.} \textbf{132} (2006) 117. [A
description od the sample of the highest-redshift quasars].}

\Ref  { Fan X., Carilli C.L. and Keating B. (2006). Observational Constraints on Cosmic Reionization. {\it Ann. Rev. Astron. \& Astrophys.} \textbf{44} 415.
[A review on the constraints drawn from quasar spectra about reionization].}

\Ref { Field G.B. (1958). Excitation of the Hydrogen 21 cm Line. {\it
Proc. IRE} \textbf{46} 240. [A classic paper on the physics of the
21 cm line of intergalactic hydrogen].}

\Ref { Field G.B. (1959).  The Time Relaxation of a Resonance-Line
Profile. {\it Astrophys. J.} \textbf{129} 551. [A pioneering paper on the
physics of the 21 cm line from intergalactic hydrogen].}

\Ref { Fukugita M. and Kawasaki M. (1994). Reionization during Hierarchical
Clustering in a Universe Dominated by Cold Dark Matter. {\it
Mon. Not. Roy. Astron. Soc.} \textbf{269} 563. [An early discussion on
cosmic reionization, about a decade before the topic gained popularity].}

\Ref  { Furlanetto S.R. and Loeb A. (2003). Metal Absorption Lines as Probes of the Intergalactic Medium Prior to the Reionization Epoch. {\it Astrophys. J.}\textbf{588} 18. [A discussion on the detectability of absorption 
lines from heavy elements which were produced in stellar interiors
and then dispersed into intergalactic space]. }

\Ref { Furlanetto S.R., Zaldarriaga M. and Hernquist L. (2004). The Growth
of H II Regions During Reionization. {\it Astrophys. J.} \textbf{613} 1.
[A calculation of the size distribution of ionized bubbles during the epoch
of reionization].}

\Ref  { Furlanetto S.R., Oh S.P. and Briggs F. (2006). Cosmology at low frequencies: The 21 cm transition and the high-redshift Universe. {\it Phys. Rep.} \textbf{433} 181. [An overview on 21cm cosmology].}

\Ref { Gehrels N. et al. (2004). The Swift Gamma-Ray Burst Mission. {\it
Astrophys. J.} \textbf{611} 1005. [A description of the SWIFT satellite
that is currently detecting gamma-ray bursts and their afterglows].}

\Ref { Gnedin N.Y. and Ostriker J.P. (1997). Reionization of the Universe
and the Early Production of Metals. {\it Astrophys. J.}  \textbf{486} 581.
[An early numerical simulation of the production of heavy elements during
the epoch of reionization]. }

\Ref { Gnedin N.Y. and Hui L. (1998).  Probing the Universe with the
Lyman-alpha forest - I. Hydrodynamics of the low-density intergalactic
medium. {\it Mon. Not. Roy Astron. Soc.} \textbf{296} 44. [A simple model
for the Lyman-$\alpha$ forest in quasar spectra].}

\Ref { Gnedin N.Y. (2000). Effect of Reionization on Structure Formation in
the Universe.  {\it Astrophys. J.} \textbf{542} 535. [Results from early
simulations of the effect of reionization on the assembly
of gas into low-mass galaxies].}

\Ref { Goodman J. (1995). Geocentrism reexamined. {\it Phys. Rev. D}
\textbf{52} 1821. [A discussion on existing evidence for the homogeneity of
the Universe].}

\Ref { Gunn J.E. and Peterson B.A. (1965). On the Density of Neutral
Hydrogen in Intergalactic Space.  {\it Astrophys. J.} \textbf{142} 1633.
[A seminal paper on the Lyman-$\alpha$ absorption feature of intergalactic
hydrogen].  }

\Ref { Haiman Z., Thoul A.A. and Loeb A. (1996). Cosmological Formation of
Low-Mass Objects. {\it Astrophys. J.} \textbf{464} 52. [The first
(spherically-symmetric) simulation of the formation of the first gas-rich
galaxies].}

\Ref { Haiman Z. and Loeb A. (1997). Signatures of Stellar Reionization of
the Universe. {\it Astrophys. J.} \textbf{483} 21. [An early detailed
calculation of reionization by stars in the modern context of cosmological
structure formation].}

\Ref { Haiman Z., Rees M.J., Loeb A. (1997).  Destruction of Molecular
Hydrogen during Cosmological Reionization. {\it Astrophys. J.} \textbf{476}
458; erratum -- {\it Astrophys. J.} \textbf{484} 985. [Negative feedback of
UV photons on the production of molecular hydrogen in the first
galaxies]. }

\Ref { Haislip J. et al. (2006). A photometric redshift of $z = 6.39 \pm
0.12$ for GRB 050904. {\it Nature} \textbf{440} 181.  [The discovery of a
gamma-ray burst with the highest redshift known].}

\Ref { Hirata C.M. (2006). Wouthuysen-Field coupling strength and
application to high-redshift 21-cm radiation. {\it
Mon. Not. Roy. Astron. Soc.} \textbf{367} 259. [A detailed discussion on
the coupling between the spin temperature and the kinetic temperature of
hydrogen through its interaction with Lyman-$\alpha$ photons]. }

\Ref  { Hogan C.J. and Rees M.J. (1979). Spectral appearance of non-uniform gas at high Z. {\it Mon. Not. Roy. Astron. Soc.} \textbf{188} 791.
[A pioneering discussion on the use of resonant lines to probe the 
intergalactic gas and study cosmology].}

\Ref { Hu E.M., Cowie L.L., McMahon R.G., Capak P., Iwamuro F., Kneib J.P.,
Maihara T. and Motohara K. (2002). A Redshift z=6.56 Galaxy behind the
Cluster Abell 370. {\it Astrophys. J. Lett.} {\textbf 568} L75. [A
spectroscopic detection of one of the earliest galaxies known].  }

\Ref { Iye M. et al. (2006). A galaxy at a redshift z = 6.96.  {\it Nature}
\textbf{443} 186. [A spectroscopic detection of one of the earliest
galaxies known]. }

\Ref { Kaiser N. (1984). On the spatial correlations of Abell
clusters. {\it Astrophys. J. Lett.} \textbf{284} L9. [A pioneering
discussion on the concept of bias in the clustering statistics of
cosmological objects].}

\Ref { Kaiser N. (1987). Clustering in real space and in redshift
space. {\it Mon. Not. Roy. Astron. Soc.} \textbf{227} 1. [A pioneering
discussion on the effect of peculiar velocities on the clustering of
sources in redshift surveys].}

\Ref { Kamionkowski M., Spergel D.N. and Sugiyama N. (1994). Small-scale
cosmic microwave background anisotropies as probe of the geometry of the
universe. {\it Astrophys. J. Lett.} \textbf{426} L57. [An early discussion
on the use of microwave background data to constrain the underlying
geometry of the Universe].}

\Ref { Kitayama T. and Ikeuchi S. (2000). Formation of Subgalactic Clouds
under Ultraviolet Background Radiation. {\it Astrophys. J.}  \textbf{529}
615. [Spherically-symmetric simulations of the suppressing effect of UV
radiation on the collapse low-mass gas coulds].}

\Ref { Kolb E.W. and Turner M.S. (1990). The early universe. (Redwood City,
CA: Addison-Wesley). [A textbook on the interface 
between modern cosmology and particle physics].}

\Ref { Komatsu E. et al. (2008). Five-Year Wilkinson Microwave Anisotropy
Probe (WMAP) Observations: Cosmological Interpretation, ArXiv e-prints,
803, arXiv:0803.0547. [The latest cosmological constraints based
on five-years of data gathering by the WMAP satellite].}

\Ref { Lamb D.Q and Reichart D.E. (2000). Gamma-Ray Bursts as a Probe of
the Very High Redshift Universe. {\it Astrophys. J.} \textbf{536} 1.  [An
early discussion on the detectability of gamma-ray bursts out to very high
redshifts].}

\Ref { Lidz A., Oh S.P. and Furlanetto S.R. (2006). Have We Detected Patchy
Reionization in Quasar Spectra? {\it Astrophys. J. Lett.} \textbf{639}
L47. [An analysis of the implications from absorption spectra of
high-redshift quasar for models of reionization]. }

\Ref { Loeb A. (2008). First Light.  {\it SAAS-Fee Advanced Course 36},
Springer Verlag, Berlin 2008. http://arxiv.org/abs/astro-ph/0701024.  [A
recent overview of the underlying physics in studies of the first
galaxies]. }

\Ref  { Loeb A. (2006). The dark ages of the Universe. {\it Scientific American}, \textbf{295}, 46 (http://www.cfa.harvard.edu/~loeb/sciam.pdf).
[A popular level review on the first galaxies and 21-cm cosmology].}

\Ref { Loeb A. and Rybicki G. (1999). Scattered Lyman-$\alpha$ Radiation
around Sources before Cosmological Reionization {\it Astrophys. J.}
\textbf{524}, 527. [A derivation of the halo of scattered
Lyman-$\alpha$ photons around a source embedded in an expanding
intergalactic medium].}

\Ref { Loeb A. and Zaldarriaga M. (2004). Measuring the Small-Scale Power
Spectrum of Cosmic Density Fluctuations through 21cm Tomography Prior to
the Epoch of Structure Formation. {\it Phys. Rev. Lett.} \textbf{92}
211301.  [The first calculation of the power-spectrum of 21-cm brightness
fluctuations during the dark ages [prior to the appearance of the first
galaxies)].}

\Ref { Loeb A. and Wyithe S. (2008). Precise Measurement of the
Cosmological Power Spectrum With a Dedicated 21cm Survey After
Reionization.  {\it Phys. Rev. Lett.} \textbf{in press}, ArXiv e-prints,
801, arXiv:0801.1677. [A study demonstrating that future 21cm data
after reionization can map the matter distribution through most of the
observable volume of the Universe].}

\Ref  { Ma C. and Bertschinger E. (1995). Cosmological Perturbation Theory in the Synchronous and Conformal Newtonian Gauges. {\it Astrophys. J.} \textbf{455} 7. [A comprehensive discussion on the growth of structure
in the Universe].}

\Ref { Madau P., Meiksin A. and Rees M.J. (1997). 21 Centimeter Tomography
of the Intergalactic Medium at High Redshift. {\it Astrophys. J.}
\textbf{475} 429. [An early discussion on the use of the 21cm line for
three-dimensional mapping of intergalactic hydrogen].}

\Ref { McQuinn M., Zahn O., Zaldarriaga M., Hernquist L. and Furlanetto
S.R. (2006). Cosmological Parameter Estimation Using 21 cm Radiation from
the Epoch of Reionization. {\it Astrophys. J.} \textbf{653} 815.  [A
demonstration of the power of statistical analysis of future 21cm data for
constraining cosmological parameters].}

\Ref { Mellema G., Iliev I.T., Pen U.L. and Shapiro P.R. (2006). Simulating
Cosmic Reionization at Large Scales II: the 21-cm Emission Features and
Statistical Signals.  {\it Mon. Not. Roy. Astron. Soc.} \textbf{372} 679.
[Results from a numerical simulation of reionization by the 
first galaxies].}


\Ref { Miralda-Escud\'e J. (1998). Reionization of the Intergalactic Medium
and the Damping Wing of the Gunn-Peterson Trough. {\it Astrophys. J.}
\textbf{501} 15. [A derivation of the spectral profile of Lyman-$\alpha$
absorption by a neutral intergalactic medium around a high-redshift
source].}

\Ref { Miralda-Escud\'e J. and Rees M J (1998). Searching for the Earliest
Galaxies Using the Gunn-Peterson Trough and the Lyman-alpha Emission
Line. {\it Astrophys. J.} \textbf{497} 21. [An early discussion on the
spectral signatures of high-redshift galaxies].}

\Ref { Miralda-Escud\'e J. (2000). Soft X-Ray Absorption by High-Redshift
Intergalactic Helium. {\it Astrophys. J. Lett.} \textbf{528} L1.  [A
discussion on the absorption signature of a neutral intergalactic medium
around an X-ray source]. }

\Ref { Murray N., Quataert E. and Thompson T.A.  (2005). On the Maximum
Luminosity of Galaxies and Their Central Black Holes: Feedback from
Momentum-driven Winds. {\it Astrophys. J.} \textbf{618} 569. [A model for
momentum-regulated growth of supermassive black holes in galaxies].}

\Ref { Naoz S. and Barkana R. (2005). Growth of linear perturbations before
the era of the first galaxies. {\it Mon. Not. Roy. Astron. Soc.}
\textbf{362} 1047. [A precise calculation of the linear evolution of
density and temperature fluctuations in the cosmic gas during the dark
ages].}

\Ref{ Navarro J. F., Frenk C. S., White S. D. M. (1997). A Universal
Density Profile from Hierarchical Clustering. {\it Astrophysical Journal}
490, 493.  [Results from numerical simulations that demonstrated the
existence of a universal form for the density profile in dark matter
halos].  }

\Ref { Navarro J.F. and Steinmetz M. (1997). The Effects of a Photoionizing
Ultraviolet Background on the Formation of Disk Galaxies. {\it
Astrophys. J.}  \textbf{478} 13. [Results from three-dimensional
simulations on the effect of UV radiation on the
assembly of gas in low-mass galaxies].}

\Ref { Oh S. P. (2001). Reionization By Hard Photons. I. X-Rays From the
First Star Clusters. {\it Astrophys. J.} \textbf{553} 499. 
[A discussion on reionization by X-ray photons].}

\Ref  { Osterbrock D.E. (1974). Astrophysics of gaseous nebulae. (San Francisco: W. H. Freeman and Company) p.14. [A textbook describing the physics 
of ionized regions around a UV source].}

\Ref { Peebles P.J.E. (1980). The large-scale structure of the universe.
(Princeton: Princeton University Press). [A textbook describing basic
concepts related to the growth of structure in the Universe].}

\Ref  { Peebles P.J.E. (1984). Dark matter and the origin of galaxies and globular star clusters. {\it Astrophys. J.} \textbf{277} 470. 
[A pioneering discussion on cold dark matter in galaxies].}

\Ref { Peebles P.J.E. (1993).  Principles of physical
cosmology. (Princeton: Princeton University Press). [A textbook on basic
concepts in the physics of cosmology].}

\Ref { Peebles P.J.E. and Yu J.T. (1970). Primeval Adiabatic Perturbation
in an Expanding Universe. {\it Astrophys. J.} \textbf{162} 815.  [A
pioneering calculation of the temperature anisotropies of the microwave
background].}

\Ref { Pritchard J.R. and Furlanetto S.R. (2006) Descending from on high:
Lyman-series cascades and spin-kinetic temperature coupling in the 21-cm
line. {\it Mon. Not. Roy. Astron. Soc.} \textbf{367} 1057.  [A
comprehensive discussion on the effect of Lyman-series photons on the spin
temperature of hydrogen and the corresponding 21-cm fluctuations].}

\Ref { Pritchard J.R. and Furlanetto S.R. (2007). 21 cm fluctuations from
inhomogeneous X-ray heating before reionization. {\it
Mon. Not. Roy. Astron. Soc.} \textbf{376} 1680. [A study of the effect of
inhomogeneous X-ray heating by the first galaxies on 21-cm fluctuations]. }

\Ref {Pritchard J.R. and Loeb A. (2008). Evolution of the 21 cm signal
throughout cosmic history.  ArXiv e-prints, 802, arXiv:0802.2102. [A
comprehensive summary of the various sources of 21-cm fluctuations at all
redshifts].}

\Ref { Purcell E.M. and Field G.B. (1956). Influence of Collisions upon
Population of Hyperfine States in Hydrogen. {\it Astrophys. J.}
\textbf{124} 542. [A pioneering paper on the effect of atomic collisions on
the spin temperature of hydrogen].}

\Ref { Quinn T., Katz N. and Efstathiou G. (1996). Photoionization and the
formation of dwarf galaxies. {\it Mon. Not. Roy Astron. Soc. Lett.}
\textbf{278} 49. [Results from simulations concerning the suppressing
effect of ionizing radiation on the assembly of gas in low-mass galaxies].}


\Ref { Rees M.J. and Sciama D.W. (1968). Larger scale Density
Inhomogeneities in the Universe. {\it Nature} \textbf{217} 511.  [A
pioneering discussion on the imprint of large scale inhomogeneities on
temperature anisotropies of the microwave background through the
time-dependence of the gravitational potential].}

\Ref { Sachs R.K. and Wolfe A.M. (1967). Perturbations of a Cosmological
Model and Angular Variations of the Microwave Background. {\it
Astrophys. J.} \textbf{147} 73. [A pioneering formal derivation of the
temperature anisotropies in the microwave background owing to density
fluctuations].}

\Ref { Scott D. and Rees M.J. (1990). The 21-cm line at high redshift: a
diagnostic for the origin of large scale structure. {\it
Mon. Not. Roy. Astron. Soc.} \textbf{247} 510. [An early discussion on the
potential use of the 21-cm line for cosmological studies].}

\Ref { Seljak U. and Zaldarriaga M. (1996). A Line-of-Sight Integration
Approach to Cosmic Microwave Background Anisotropies. {\it Astrophys. J.}
\textbf{469} 437. [An efficient simplified solution to the equations that
provide the microwave background anisotropies].}

\Ref { Shapiro P.R. and Giroux M.L. (1987). Cosmological H II regions and
the photoionization of the intergalactic medium.  {\it Astrophys. J. Lett.}
\textbf{321} L107. [An early discussion on the evolution of ionized regions
around a UV source embedded within an expanding medium of cosmic gas].}

\Ref { Shapiro P.R., Giroux M.L. and Babul A. (1994). Reionization in a
cold dark matter universe: The feedback of galaxy formation on the
intergalactic medium. {\it Astrophys. J.} \textbf{427} 25.  [An early
discussion on reionization by galaxies].}

\Ref { Silk J. (1968). Cosmic Black-Body Radiation and Galaxy
Formation. {\it Astrophys. J.} \textbf{151} 459. [A pioneering 
derivation of the effect of photon diffusion on the damping 
of microwave background anisotropies on small scales].}

\Ref { Silk J. and Rees M.J. (1998). Quasars and Galaxy Formation.  {\it
Astron. \& Astrophys.} \textbf{331} L1. [A schematic discussion on the
expected scaling relations between central black hole mass and
velocity dispersion in galaxies, based on self-regulated growth by 
momentum or energy feedback].}


\Ref { Stark D.P., Loeb A. and Ellis R. (2007). An Empirically Calibrated
Model for Interpreting the Evolution of Galaxies during the Reionization
Era. {\it Astrophys. J.} \textbf{668} 627. [A simple theoretical model 
for observations of high redshift galaxies].}

\Ref { Sunyaev R.A. and Zeldovich Y.B. (1970). Small-Scale Fluctuations of
Relic Radiation. {\it APSS} \textbf{7} 3. [A pioneering derivation of the
acoustic oscillation signature in the anisotropies of the microwave
background].}

\Ref { Tegmark M. et al. (1997). How Small Were the First Cosmological
Objects? {\it Astrophys. J.} \textbf{474} 1. [An early discussion 
on the conditions for the formation of the first galaxies]. }


\Ref { Thoul A.A. and Weinberg D.H. (1996). Hydrodynamic Simulations of
Galaxy Formation. II. Photoionization and the Formation of Low-Mass
Galaxies. {\it Astrophys. J.} \textbf{465} 608. [Results
from a spherically-symmetric simulation of the suppressed collapse
of gas clouds under the influence of a UV radiation background].}

\Ref { Totani T., Kawai N., Kosugi G., Aoki K., Yamada T., Iye M., Ohta
K. and Hattori T. (2006). Implications for Cosmic Reionization from the
Optical Afterglow Spectrum of the Gamma-Ray Burst 050904 at $z = 6.3$. {\it
Pub. Astron. Soc. Japan} \textbf{58} 485. [A discussion on the implication
from the spectral data of the gamma-ray burst with the highest known
redshift].}

\Ref { Trac H., and Cen R. (2007). Radiative Transfer Simulations of Cosmic
Reionization. I. Methodology and Initial Results.  {\it Astrophys. J.}
\textbf{671} 1. [Results from a state-of-the-art computer simulation of
cosmic reionization].}

\Ref { Verner D.A., Ferland G.J., Korista T. and Yakovlev
D.G. (1996). Atomic Data for Astrophysics. II. New Analytic FITS for
Photoionization Cross Sections of Atoms and Ions. {\it Astrophys. J.}
\textbf{465} 487. [A compilation of photo-ionization cross sections for a
variety of atomic and ionic species].}


\Ref {Weinberg S. (1972). Gravitation and Cosmology: Principles and
Applications of the General Theory of Relativity. {\it Gravitation and
Cosmology} (New York: Wiley). [A pioneering textbook that established the
currently popular link between cosmology and particle physics].}

\Ref { Weinberg D.H., Hernquist L. and Katz N. (1997). Photoionization,
Numerical Resolution, and Galaxy Formation. {\it Astrophys. J.}
\textbf{477} 8. [Results from numerical simulations on the effect of
ionizing radiation on galaxy formation].}

\Ref { White R.L., Becker R.H., Fan X., Strauss M.A. (2003). Probing the
Ionization State of the Universe at $z>6$. {\it Astron. J.} \textbf{126}
1. [A study of the inference from quasar spectra concerning the ionization
state of the intergalactic medium at redshift $z>6$].}

\Ref { Wouthuysen S.A. (1952). On the excitation mechanism of the 21-cm
(radio-frequency) interstellar hydrogen emission line. {\it Astron. J.}
\textbf{57} 31. [A pioneering study of the effect of Lyman-$\alpha$ photons
in couplin the spin temperature of hydrogen to its kinetic temperature].}

\Ref { Wu K.K.S., Lahav O. and Rees M.J. (1999). The large-scale smoothness
of the Universe. {\it Nature} \textbf{397} 225. [A summary of the evidence
for the isotropy and homogeneity of the Universe].}

\Ref { Wyithe J.S.B. and Loeb A. (2003). Self-regulated Growth of
Supermassive Black Holes in Galaxies. {\it Astrophys. J.} \textbf{595}
614. [An early model for self-regulated growth of quasars and their
resulting luminosity function].}

\Ref { Wyithe J.S.B. and Loeb A. (2004a).  A large neutral fraction of
cosmic hydrogen a billion years after the Big Bang. {\it Nature}
\textbf{427} 815. [A study linking the size of the ionized regions around
high-redshift quasars with the neutral fraction of the intergalactic
medium].}

\Ref { Wyithe J.S.B. and Loeb A (2004b).  A characteristic size of $\sim$
10Mpc for the ionized bubbles at the end of cosmic reionization. {\it
Nature} \textbf{432} 194. [A model-independent derivation of the
characteristic size of ionized bubbles at the end of the reionization
epoch].}

\Ref { Wyithe J.S.B., Loeb A. and Barnes D.G. (2005). Prospects for
Redshifted 21 cm Observations of Quasar H II Regions. {\it Astrophys. J.}
\textbf{634} 715. [A study of the feasibility of detecting ionized regions
around high-redshift quasars as cavities in 21-cm surveys].}

\Ref { Wyithe J.S.B., and Loeb A. (2008). Fluctuations in 21-cm emission
after reionization. {\it Mon. Not. Roy. Astron. Soc.}, \textbf{383}, 606.
[A derivation of 21-cm fluctuations at low redshifts owing to dense
(Galactic) pockets of hydrogen that are self-shielded from the UV
background after reionization].}

\Ref { Wyithe J.S.B., Loeb A. and Geil P.M. (2008). Baryonic acoustic
oscillations in 21-cm emission: a probe of dark energy out to high
redshifts. {\it Mon. Not. Roy. Astron. Soc.} \textbf{383}, 1195. [A study
of the detectability of baryonic acoustic oscillations in the 21-cm
brightness fluctuations after reionization].}

\Ref { Yamamoto K., Sugiyama N. and Sato H. (1997). Cosmological baryon
sound waves coupled with the primeval radiation. {\it Phys. Rev. D}
\textbf{56} 7566. [A comprehensive discussion on the underlying physics of
acoustic oscillations]. }


\Ref { Yoshida N., Omukai K., Hernquist L., and Abel T. (2006).  Formation
of Primordial Stars in a LCDM Universe. {\it Astrophys. J.} \textbf{652}
6. [Simulations of the formation of the first stars over a large
cosmological region]. }

\Ref { Zahn O., Lidz A., McQuinn M., Dutta S., Hernquist L., Zaldarriaga
M. and Furlanetto S.R. (2006). Simulations and Analytic Calculations of
Bubble Growth During Hydrogen Reionization. {\it Astrophys. J.}
\textbf{654} 12. [Results from simulations of 
the evolution of cosmic reionization].}

\Ref { Zhang W., Woosley S. and MacFadyen A.I. (2003). Relativistic Jets in
Collapsars. {\it Astrophys. J.} \textbf{586} 356. [Simulations of the
popular collapsar model for long-duration gamma-ray bursts, in which
relativistic jets are produced by the collapse of a stellar core to a black
hole]. }

\Ref { Zygelman B. (2005). Hyperfine Level-changing Collisions of Hydrogen
Atoms and Tomography of the Dark Age Universe. {\it Astrophys. J.}
\textbf{622} 1356. [A detailed discussion on the effect of atomic collisions
on the spin temperature of cosmic hydrogen].}

\vfill\eject

\end{document}